\documentclass[preprint,authoryear,12pt]{elsarticle}

\usepackage{amsbsy}
\usepackage{hyperref}

\usepackage{booktabs}

\usepackage{amsmath}

\usepackage{graphicx}

\graphicspath{{./Boxplots_Apr2019/}}

\usepackage{subfigure}
\usepackage{multirow}
\usepackage{url}

\newcommand{\blind}{0}

\newcommand{\be}{\begin{equation}}
\newcommand{\ee}{\end{equation}}

\newcommand{\bbeta}{{\boldsymbol{\beta}}}
\newcommand{\btheta}{{\boldsymbol{\theta}}}

\newcommand{\U}{\boldsymbol{U}}

\renewcommand{\u}{\boldsymbol{u}}

\usepackage[usenames]{color}
\usepackage[normalem]{ulem}

\journal{Communications in Statistics: Simulation and Computation}

\begin{document}

\begin{frontmatter}

  \title{Copula Density Estimation by Finite Mixture of Parametric Copula Densities}
  \author{Leming Qu\corref{Qu}}
  \author{Yang Lu\corref{Lu}}
  \cortext[Qu]{Corresponding author. Leming Qu is with Department of Mathematics, Yang Lu is with Department of Civil Engineering, Boise State University, Boise, Idaho 83725, USA. Emails: lqu@boisestate.edu; yanglufrank@boisestate.edu}


\begin{abstract}
A Copula density estimation method that is based on 
a finite mixture of heterogeneous parametric copula densities is proposed here. More specifically, the mixture components are Clayton, Frank,  Gumbel, T, and normal copula densities,  which are capable of capturing lower tail, 
 strong central, upper tail, heavy tail, and symmetrical elliptical dependence, respectively. 
The model parameters are estimated by an interior-point algorithm for the constrained maximum likelihood problem.
The interior-point algorithm is compared with the commonly used EM algorithm. 
Simulation and real data application show that the proposed approach is effective to model complex dependencies for data in dimensions beyond two or three.      



\end{abstract}


\end{frontmatter}





\section{Introduction}

Dependence modeling consists of finding a model that describes dependencies between 
variables, which is a fundamental task of multivariate statistics (\cite{CoxWermuth1996}).  
A statistical approach to dependence modeling describes an underlying random process 
in terms of a multivariate distribution. Multivariate probability density estimation
based on observed data from a random process is a long standing and active research 
area in statistics (\cite{Scott1992}).  In a linear, Gaussian world stochastic dependencies
are captured by correlations. In more general settings, copula (otherwise known as 
dependence function) has emerged as a
useful tool for modeling stochastic dependence (\cite{Joe2015, HofertEtal_Springer_2018}). 
In essence, a copula is a multivariate probability distribution with uniform marginals. 
One of the main advantages of a copula over a full probability function is that a 
copula allows the separation of dependence modeling from the marginal distributions. 

The copula density estimation can be categorized into parametric, semiparametric, and nonparametric methods. 
A parametric estimation method assumes both the copula density and all the marginal densities belong to some parametric families determined by a few parameters (for example, \cite{ShihLouis1995}).
The parametric copula density estimation problem is then essentially reduced to estimate the few parameters
that determine the copula and the marginal densities. 

Nonparametric estimation of a copula density does not assume a specific parametric form
for the copula density and thus provides great flexibility and generality. For example, 
\cite{Racine2015} proposed a kernel-based copula density estimator and provided an R package \texttt{np} (\cite{HayfieldRacine_JSS_Jul2008}).  
\cite{KauermannSchellhaseRuppert_SJS_Dec13_P685} fitted a copula density using 
penalized hierarchical B-splines in sparse grids and implemented it in an R package \texttt{pencopula}. See the Introduction section of \cite{KauermannSchellhaseRuppert_SJS_Dec13_P685} for a brief review of nonparametric copula density estimation literature.  

Semiparametric copula density estimation method assumes part of the data distribution - such as the copula density - follows a parametric model, 
while the rest - such as the univariate marginal distributions - follow nonparametric models. The two stage estimation method (\cite{GenestGhoudiRivest_Biometrika_Sep95_P543}) for iid data proceeds as following: (1) in the first stage, an univariate marginal distribution is estimated nonparametrically, 
e.g., by the rescaled empirical marginal distribution; (2) in
the second stage, the copula parameters are estimated by maximizing the pseudo log-likelihood using the data 
generated in the first stage. The resulting semi-parametric estimator of the dependence
parameter is consistent and asymptotically normal under suitable regularity conditions. 
The two stage estimator for iid data has been extended to time series setting (\cite{ChenFan_JoEM_Feb2006_P307, ChenFan_JoEM_Nov2006_P125}).
\cite{ChenEtal_JASA_2006}  propose a sieve maximum likelihood estimation procedure which is semiparametrically
efficient. 

We propose here to estimate a multivariate copula density by a finite mixture of heterogeneous parametric copulas, which further enhance the 
flexibility of multivariate distribution modeling. 
Mixture probability density function comprising a 
finite number of components, possibly of different types of probability density that can capture diverse features in the data, 
offers a less restrictive parametric modeling  as an interesting alternative to nonparametric modeling. 
Finite mixture models are widely used in statistical data analysis and there exist extensive literature on this modeling framework - see, for example, the books by \cite{TitteringtonEtal1985, Lindsay1995, Bohning1999, MclachlanPeel2004, Fruhwirth2006, MengersenEtal2011}.  


For the copula density estimation problem, there are several papers which use a finite mixture of parametric copula densities modeling approach.  
\cite{Hu2006} uses a mixture of three copulas to capture various symmetric and asymmetric dependence
structures in financial markets. The mixture is composed of a Gaussian copula, a Gumbel
copula and a Gumbel survival copula. The Gaussian copula in the mixture relates to traditional approaches based on the Gaussian
assumption. Gumbel copula and its survival copula model extreme co-movements in market returns.
The former models positive right tail dependence while the latter is its mirror image and models left tail dependence. 
In \cite{Hu2006},  the mixture model is estimated by a two-stage semi-parametric procedure, i.e. the marginals are estimated by the empirical distributions. EM algorithm is then used to maximize the pseudo log-likelihood. \cite{Hu2006} considers only bivariate copulas.
\cite{KauermannMeyer2014} proposes a finite mixture of different Archimedean copula families 
as a flexible tool for modeling the dependence structure in multivariate data.
The parameters in this mixture model are estimated by maximizing
the penalized marginal likelihood via iterative quadratic programming. 
A fully Bayesian approach via simulation-based posterior computation is also presented.
\cite{KauermannMeyer2014} considers only Archimedean copula families. 
\cite{ArakelianKarlis2014} uses a finite mixture of different copulas for clustering purposes, with parametric marginal distributions. 
The model parameters are estimated by an EM algorithm based on the standard approach for mixture models. \cite{ArakelianKarlis2014} focuses on bivariate models. \cite{CaiWang2014} selects an appropriate mixed copula and estimates the related parameters simultaneously via penalized likelihood plus a shrinkage operator. The EM algorithm is used to find the penalized likelihood estimator and a data-driven method is used
to find the tuning and thresholding parameters in the penalty function. The simulated examples and real data analysis in \cite{CaiWang2014} 
are applied to bivariate data sets.  

There are very few papers which use an infinite mixture of parametric copula densities modeling approach. \cite{WuWangWalker_JSCS_85(1)_2015_P103} shows that any bivariate copula density can be arbitrarily accurately approximated by an infinite mixture of Gaussian copula density functions and that the model can be estimated by
	Markov Chain Monte Carlo (MCMC) methods. \cite{WuWangWalker_MCAP_Sep14_P747} constructs a nonparametric copula density by an infinite mixture of
	multivariate skew–normal copulas and develops an MCMC algorithm to draw
	samples from the correct posterior distribution.



The main contribution of this article is to shed insight on the interior point algorithm as an useful alternative to the commonly 
	used EM algorithm for a mixture model parameter estimation. In the context of mixture copula modeling for dimensions beyond two or three, the interior point algorithm is able to fit the model well as shown both in simulation studies and in real data applications.

We mention a few papers that apply interior point algorithm to solve statistical model parameter estimation problems here. 
\cite{KoenkerPark1996}  describes an interior point algorithm for nonlinear quantile regression. 
\cite{KoenkerMizera2014a} reformulates the Kiefer-Wolfowitz nonparametric maximum likelihood estimator for mixtures as a convex optimization problem. 
\cite{KimEtal2007} and \cite{KohEtal2007} apply an interior point method for large-scale l1-regularized least squares and logistic regression problem, respectively.

The rest of the paper is organized as follows. 
In section \ref{sec:model}, we present the finite mixture of parametric copulas model.
In section \ref{sec:alg}, we discuss the interior point algorithm and compare it with the classical expectation-maximization (EM) algorithm. 
Section \ref{sec:simu}  shows the experimental results.
We apply the method to two real data sets in section \ref{sec:real}.
Finally, section \ref{sec:con} concludes the paper.

\section{A Finite Mixture of Heterogeneous Parametric Copulas Model} \label{sec:model}

A multivariate copula density $c(\u)$, ~$\u=(u_1, \ldots, u_p)\in [0,1]^p$  can be regarded as the joint probability density function (PDF) of
a $p$-standard uniform random variable $\U = (U_1, \ldots, U_p)$. 

A multivariate copula $C(u_1, \ldots, u_p)$  defined on a unit hypercube $[0, 1]^p$ is a $p$-variate cumulative
distribution function (CDF) with univariate standard uniform margins:
\[
C(u_1,\ldots, u_p) = \int_0^{u_1}\cdots\int_0^{u_p} c(v_1,\ldots, v_p) \mathrm{d}v_1 \cdots \mathrm{d}v_p.
\]
Sklar's Theorem (\cite{Sklar1959}) states that the joint CDF $F(x_1, \ldots, x_p)$ of a $p$-variate random variable $(X_1, \ldots, X_p)$
with marginal CDF $F_j(x_j)$ can be written as 
\[
F(x_1, \ldots, x_p) = C(F_1(x_1), \ldots, F_p(x_p)),
\]
 where the copula $C$ is the
joint CDF of $(U_1, \ldots, U_p)=(F_1(X_1), \ldots, F_p(X_p))$.
This indicates a copula connects the marginal
distributions to the joint distribution and justifies the use of copulas for building multivariate distributions.

Let $(x_{11}, \ldots, x_{1p}), \ldots, (x_{n1}, \ldots, x_{np})$  be a random sample from the unknown distribution $F$ of
$(X_1, \ldots, X_p)$.  We wish to estimate aspects of the joint distribution of $X_1, \ldots, X_p$,
in particular, the copula density function $c(\u)$.

When the marginal distributions
are continuous, the copula density $c(\u)$ is the unique $p$-variate density
of $(U_1, \ldots, U_p)$ as implied by the Sklar's theorem.
As copulas are not directly observable, a copula density estimator is usually 
formed in two stages: obtaining the observations for $(U_1, \ldots, U_p)$ first and then estimating
the copula density based on these observations.

In the first stage, the original data set $(x_{i1}, \ldots, x_{ip})$ for $i=1, \ldots, n$ is converted to
$(\hat u_{i1},\ldots,\hat u_{ip}) = (\hat F_1(x_{1i}), \ldots, \hat F_p(x_{ip}))$, where $\hat F_1(\cdot), \ldots, \hat F_p(\cdot)$ are
conventional estimators of $F_1(\cdot), \ldots, F_p(\cdot)$. If a parametric model, such as a T-distribution, is appropriate for a marginal 
distribution $F_j(\cdot)$, one can use a technique such as maximum likelihood method
to estimate its parameters.  Otherwise, some nonparametric univariate distribution
estimation methods or simply the empirical CDF can be used.

In the second stage, we estimate the copula density $c(\u)$ based on the observations
$\{(\hat u_{i1}, \ldots, \hat u_{ip})\}_{i=1}^n$.

We assume the copula density $c(\u)$ a finite mixture of five different types of copula families: 
\begin{align}\label{model}
c(\u) = w_C c_C(\u;\alpha_C) + w_F c_F(\u;\alpha_F) + w_G c_G(\u;\alpha_G) + w_T c_T(\u;R_T, \nu) +
\sum_{j=1}^k w_j c_N(\u;R_j), \nonumber \\
\end{align} 
where $w_C, w_F, w_G, w_T, w_j$ denote the proportions; 
$c_C(\cdot), c_F(\cdot), c_G(\cdot), c_T(\cdot), c_N(\cdot)$ denote the densities;
and   $\alpha_C, \alpha_F, \alpha_G, (R_T, \nu), R_j$ denote the parameters of  Clayton, Frank, Gumbel, T, and normal copula respectively. 
There are $k$ normal copula components.  Mixture proportions are nonnegative and sum to one. Copula parameters are restricted within their respective parameter spaces.
For Clayton copula parameter: $\alpha_C>0$. For Frank copula parameter: $\alpha_F>0$.  For Gumbel copula parameter: $\alpha_G\ge 1$.
For T-copula parameter, $R_T$ is a $p\times p$ correlation matrix, and $\nu>0$ is its degrees of freedom. For the $j$th normal copula, $R_j$ is a $p\times p$ correlation matrix.
A correlation matrix has diagonal elements 1 and off-diagonal elements in the range $[-1, 1]$. It must be symmetric and  positive semi-definite.     

We call model (\ref{model}) a CFGTN model. One parameter Clayton, Frank, Gumbel copulas are members of the Archimedean family (\cite{Nelsen2006}, pp. 116).  Archimedean copulas are exchangeable, that is, stays the same by permutations of $u_1, \ldots, u_p$. A Clayton copula can capture lower tail  dependence. 
A Frank copula can capture strong dependence in the center of the distribution, but not tail dependence. 
A Gumbel copula can capture upper tail dependence. 

T-copula and normal copulas are members of elliptical copulas, i.e., copulas of elliptical distributions. A T-copula can capture 
symmetrical and heavy tail dependence. A normal copula can capture symmetrical dependence, but not tail dependence. The number of normal copula components $k$ is to be determined by the data using a model selection criteria such as AICc.  Using data adaptive $k$ normal copula components instead of a single one is intended to capture more complex dependence structures. 
By mixing normal copulas with a T-copula and commonly used Archimedean copulas, we believe that higher flexibility can be achieved than mixing normal copulas only. Our simulation and real data examples support that 
a mixture of Clayton, Frank, Gumbel, T, and normal copulas is capable of capturing 
most of the possible dependence structures. 

The correlation matrix $R_T$ of the T-copula has $(p-1)p/2$ unknown parameters, so does each correlation matrix $R_j$ of the normal copula. For a T-copula or a normal copula alone, the sample correlation matrix is a natural estimate of the population correlation matrix. But for a mixture model comprising a T-copula and $k$ normal copulas, it is not an easy task to estimate the $(1+k)$ many correlation matrices. Moreover, the number of unknown parameters in each correlation matrix keeps growing quadratically with dimension $p$. In this paper, we only deal with moderate dimension $p$ beyond 2 or 3. 
High-dimensional correlation matrix estimation problem alone is an active current research area [\cite{ZhaoRoederLiu_JCGS_Oct2014_P895}]. 
One of the major obstacles in correlation matrix estimation is to ensure its positive semi-definiteness.
Hyperspherical reparameterization of a correlation matrix's Cholesky factor has emerged as a flexible and effective solution
[\cite{PinheiroBates_StatComp_Sep1996_P289, RebonatoJackel_JoRisk_2(2)_Winter1999_P17, RapisardaBrigoMercurio_IMAJMM_Jan2007_P55, PourahmadiWang_SPL_Nov2015_P5, TsayPourahmadi_Biometrika_Mar2017_P237}]. 
Most recently, \cite{Yoshiba_JSCS_Sep2018_P2489} uses this reparameterization for maximum likelihood estimation of skew-t copulas.
\cite{PourahmadiWang_SPL_Nov2015_P5} summarizes the origins of this method: 
\medskip
\\
\emph{``The idea of reparameterizing the Cholesky factor of a covariance matrix using the hyperspherical coordinates is due
	to \cite{PinheiroBates_StatComp_Sep1996_P289} section 2.3. For correlation matrices, an early and naive version was proposed by \cite{RebonatoJackel_JoRisk_2(2)_Winter1999_P17}, however, \cite{RapisardaBrigoMercurio_IMAJMM_Jan2007_P55}
	 develop a more complete setup with the full geometrical implications of the idea."}
\medskip
\\
  For a normal or a multivariate Student-t
distribution model, the consistency and asymptotic normality of the maximum likelihood estimators of the  hyperspherical coordinates 
or angles for a structured correlation matrix were established in \cite{TsayPourahmadi_Biometrika_Mar2017_P237}.


The well-known Cholesky factorization of the correlation matrix $R=(r_{ij})$ of a T-copula or a normal copula is $R = LL^T$, where $L=(l_{ij})$ is a lower triangular matrix along
with its  hyperspherical  reparameterization as
\begin{equation*}
L = \left[
\begin{array}{ccccc}
l_{11} & 0      & 0 & \ldots & 0 \\
l_{21} & l_{22} & 0 & \ldots & 0 \\
l_{31} & l_{32} & l_{33} & \ldots & 0 \\
l_{41} & l_{42} & l_{43} & \ldots & 0 \\
\vdots & \vdots & \vdots & \ldots & \vdots \\
l_{p1} & l_{p2} & l_{p3} & \ldots & l_{pp} 
\end{array} \right] = \left[
\begin{array}{ccccc}
1 & 0      & 0 & \ldots & 0 \\
\cos\theta_{21} & \sin\theta_{21}          &           0              & \ldots & 0 \\
\cos\theta_{31} & \cos\theta_{32}\sin\theta_{31} & \sin\theta_{32}\sin\theta_{31} & \ldots & 0 \\
\cos\theta_{41} & \cos\theta_{42}\sin\theta_{41} & \cos\theta_{43}\sin\theta_{42}\sin\theta_{41} & \ldots & 0 \\
\vdots & \vdots & \vdots & \ldots & \vdots \\
\cos\theta_{p1} & \cos\theta_{p2}\sin_{p2} & \cos\theta_{p3}\sin\theta_{p2}\sin\theta_{p1} & \ldots & \displaystyle{\prod_{k=1}^{p-1}}\sin\theta_{pk} 
\end{array} \right]. 
\end{equation*}
Because the $R$ is a correlation matrix, we have $l_{ii}=1$ and $l_{ij} \in [-1,1]$ for $i>j$, which can be represented by  
angles $\theta_{ij}$ measured in radians for $i>j$. The angles are required to be restricted to the range $(0, \pi)$ so that the $R$ 
has positive diagonal entries and hence the Cholesky factor $L$ is unique. 
According to Lemma 1 of \cite{PourahmadiWang_SPL_Nov2015_P5},
the transformation from $R$ to $\Theta = (\theta_{ij})$ is one-to-one, where $\theta_{ij}=0$ for $i<=j$. 

On the issue of model identifiability, there are rare cases that the model is unidentifiable. 
One such case is when two or more mixture components have the product copula
$C(u_1, \ldots, u_p) = u_1\cdots u_p$ as a special case for specific values of their
parameters (for example, the Clayton copula for $\alpha_C \rightarrow 0$,
the Frank for $\alpha_F \rightarrow 0$,  the Gumbel for $\alpha_G \rightarrow 1$, 
the normal copula for the correlation matrix $R$ approaching to identity $I$,  and so on).
For Archimedean copulas, these cases happen when the copula parameters are near specific boundary 
values of their respective parameter space [\cite{KosmidisKarlis_StatComp_Sep2016_P1079}]. 
However, if the main interest is to estimate the mixture density $c(\u)$ rather than to identify the individual mixture component as a cluster, 
the density estimate itself $\hat c(\u)$ is unaffected by the label switching problem, since it does not depend on how the 
components are labeled [\cite{Stephens_JRSSB_62(4)_2000_P795}].

It is well known that label switching results in difficulties for finite mixture models and 
simple inequality constraints on the parameter space can be used to break 
the symmetry in the likelihood [\cite{RichardsonGreen_JRSSB_59(4)_1997_P731}, \cite{JasraHolmesStephens_STS_Feb05_P50}].
For normal copulas, it is more natural and easier to impose simple inequality constraints on the scalar mixture proportions than to 
 impose some constraints on the $p\times p$ dimensional correlation matrices.  
 We therefore put the normal copula proportions $w_1, \ldots, w_k$ in non-increasing order in the model specification. 

The maximum pseudo log-likelihood estimator $\hat\bbeta$ in constrained parameter spaces maximizes the pseudo log-likelihood
\begin{equation}
\begin{aligned} \label{mple}
 \max_\bbeta L(\bbeta) & = \sum_{i=1}^n \log c(\hat \u_i)  \\
          & = \sum_{i=1}^n \log [w_C c_C(\hat\u_i;\alpha_C) + w_F c_F(\hat\u_i;\alpha_F) + w_G c_G(\hat\u_i;\alpha_G)  \\
          & ~~~ + w_T c_T(\hat\u_i;R_T, \nu) + \sum_{j=1}^k w_j c_N(\hat \u_i; R_j)] ,  \\
           \mbox{Subject to}  \\
          & w_C\ge 0,  ~w_F\ge 0, ~w_G\ge 0, ~w_T\ge 0;  ~ w_1\ge \ldots \ge w_k \ge 0 \\
          & w_C + w_F + w_G + w_T + \sum_{j=1}^k w_j = 1  \\
          & \alpha_C \ge 0, ~\alpha_G \ge 1, ~\nu>0, \\
          & \btheta_T \in (0, \pi), ~ \btheta_i \in (0, \pi), ~ \mbox{ for } i=1, \ldots, k, 
\end{aligned}
\end{equation}
where $\bbeta=(w_C, w_F, w_G, w_T, w_1, \ldots, w_k, \alpha_C, \alpha_F, \alpha_G, \nu,  \btheta_T, \btheta_1, \ldots, \btheta_k)$ is the vector of unknown proportions, copula parameters, 
and angles for the correlation matrices.  

In the next section, we discuss algorithms for solving this optimization problem.

\section{Constrained Maximum Likelihood Estimation by the Interior Point Algorithm} \label{sec:alg}

We first briefly review the interior point algorithm for  solving problems like (\ref{mple}) in this section,
then discuss some specifics when applying this algorithm to our problem (\ref{mple}).
There is a rich body of literature on this topic in mathematical programming (\cite{Wright1992, ByrdEtal1999, ByrdEtal2000, WaltzEtal2006, Wright1997}).   
Problem (\ref{mple}) is a special case of the following constrained nonlinear optimization (or programming) problem:
\begin{equation}
\begin{aligned} \label{con_non_opt}
\min_\bbeta  & L(\bbeta),\\
\mbox{Subject to }  \\
          & h(\bbeta) = 0  \\
          & g(\bbeta) \le 0,  \\
\end{aligned}
\end{equation}
where $L(\cdot): R^p \Rightarrow R$, $h(\cdot): R^p \Rightarrow R^l$ and $g(\cdot): R^p \Rightarrow R^m$ are twice continuously differentiable functions (\cite{WaltzEtal2006}). 

The interior point approach to this constrained minimization is to replace the inequality constraints by log barrier (Lagrangian) penalty functions that introduce
a smooth contribution to the objective function. This leads to the replacement of the nonlinear program (\ref{con_non_opt})
 by a sequence of  approximate barrier subproblems (\cite{MATLAB2017b}). 

For each $\mu>0$, the approximate problem to the original problem (\ref{con_non_opt}) is
\begin{equation}
\begin{aligned} \label{barrier}
\min_\bbeta  & L_\mu(\bbeta,s) \equiv L(\bbeta)-\mu\sum_{i=1}^m \ln(s_i), \\
\mbox{Subject to }  \\
          & h(\bbeta) = 0,  \\
          & g(\bbeta) + s = 0.  \\
\end{aligned}
\end{equation}
Here $s$ is a vector of slack variables and its elements $s_i$ are positive to keep $\ln(s_i)$ bounded. 
The $\mu>0$ is the barrier parameter. By judicious choice of a sequence of $\mu$ decreasing to zero, the minimum of $L_\mu(\cdot)$ should approach the minimum of $L(\cdot)$.

The barrier problem (\ref{barrier}) is a sequence of equality constrained problems which are easier to solve than the original inequality-constrained problem (\ref{con_non_opt}).

To solve the barrier problem (\ref{barrier}), the algorithm uses one of the two main types of steps at each iteration:
\begin{itemize}
\item A direct step in $(\bbeta, s)$. This step attempts to solve the KKT equations - first order optimality conditions, for the barrier problem (\ref{barrier}) via a linear approximation. 
      This is also called a Newton step.
\item A CG (conjugate gradient) step, using a trust region.
\end{itemize}
By default, the algorithm first attempts to take a direct step. If it cannot, it attempts a CG step. One case where it does not take a direct step is when the approximate problem 
is not locally convex near the current iterate.

    \subsection{Thresholding and Model Selection}

In practice, the number of normal mixture components $k$ is unknown. A model involving a single normal component model with $k=1$ is simple, while the one involving a dozen normal components where $k=12$ certainly
looks complex. To choose an appropriate normal model order $k$, we use a model selection criterion.

A model selection criterion offers a trade-off between the goodness of fit of the model and the complexity of the model. 
We choose $k$ by minimizing the corrected Akaike information criterion:
\[
\mbox{AICc}(k) = -2L(\hat\bbeta_k) + 2\mbox{DF}(k) + \frac{2\mbox{DF}(k)(2\mbox{DF}(k)+1)}{n-\mbox{DF}(k)-1}, 
\]
where $L(\hat\bbeta_k)$ is the log likelihood evaluated at the fitted $\hat\bbeta_k$ for the CFGTN model with $k$ normal components, and  DF$(k)$ is its Degrees of Freedom. 
The $L(\hat\bbeta_k)$ measures the goodness of fit of the model, while  DF($k$) measures the complexity of the model. In \cite{KauermannSchellhaseRuppert_SJS_Dec13_P685}, AICc is used to select the penalty parameter for the copula density estimation with 
penalized hierarchical B-splines.  

A mixture component with small proportion such as 0.01 implies small contribution to the dependence structure, therefore should not be included in the copula model (\cite{CaiWang2014}). 
We set the threshold for proportion at 0.01 as well due to its good performance in our simulation studies. 
Any component with its fitted proportion less than or equal to this threshold will be discarded from the model, leading to a reduction of model complexity. 
Therefore, the number of effective parameters in the model is
\begin{align*}
\mbox{DF}(k) = & 2\left[I(\hat w_C>0.01) + I(\hat w_F>0.01) + I(\hat w_G>0.01)  \right] +  \\
               & \left[2+\frac{(p-1)p}{2}\right]I(\hat w_T>0.01) +  \left[1+\frac{(p-1)p}{2}\right]\sum_{j=1}^k I(\hat w_j>0.01) -1 .
\end{align*} 
where $I(\cdot)$ is an indicator function. 
 Each kept component of Clayton, Frank, Gumbel, or normal copula whose estimated proportion is above the threshold 0.01 has 
a proportion parameter and a copula parameter, hence adding 2 to DF($k$). The T-copula's correlation matrix has $(p-1)p/2$
parameter and 1 parameter for the degrees of freedom $\nu$.   
The last term $-1$ in DF($k$) is due to the constraint that all the proportions sum to 1.  

Another well known model selection criterion BIC performs similarly in our simulation study.

Our model selection strategy starts with the best single component model, i.e., one of the Clayton, Frank, Gumbel, T, or normal copula according to the  model selection criterion. The copula parameter for each parametric copula is estimated by the maximum likelihood method.  We then proceed to the mixture copula model with $k=1$. The initial value for each proportion is the one which makes equal proportions for all the mixture components.    
The initial value of each copula parameter is its corresponding maximum likelihood estimate for the single component model without mixtures. 
For example, the initial value for the Clayton copula parameter is the maximum likelihood estimate for the Clayton copula model.  If the fitted model does not improve the model selection criterion, then the algorithm stops, and the previously selected single component model is chosen as the final model. Otherwise, the algorithm continues to the next step.
 
At the next step, the $k$ is increased by 1. The initial value for each proportion is again simply the one which assigns equal proportions for all the mixture components.  The initial value for each copula parameter is its corresponding fitted value from the previous step, except that the correlation matrix for the newly added normal copula component is initialized by an identity matrix. Once the initial values are assigned, the model is fitted by the interior point algorithm and the model selection criterion AICc is calculated for the fitted model. This procedure repeats until AICc no longer improves.

    \subsection{The Interior Point Algorithm vs the EM Algorithm}

EM algorithm has dominated the literature on maximum likelihood estimation of mixture models. 
For the problem of Kiefer-Wolfowitz nonparametric maximum likelihood estimator for mixtures,  \cite{KoenkerMizera2014} compared the  
modern interior point methods with the EM algorithm. Their experience was that modern interior point methods
are vastly superior, both in terms of accuracy and computational effort.

Here we compare the interior point algorithm with the EM Algorithm for simulated data sets in 2 dimensions. We replicate the estimation procedure for a simulated data set 100 times each with $n = 1000$ observations from a mixture of bivariate Clayton, Gumbel, and normal copula with the 
parameters specified in Table \ref{table:GCN-model}. 

\begin{table}
\caption{A mixture of Clayton, Gumbel and normal copula model for simulation}
\label{table:GCN-model}
\begin{center}
\begin{tabular} {l|l|c}
\hline
Component & Proportion & parameter \\
\hline
\hline
Clayton   &   0.40     &  3   \\
Gumbel    &   0.25     &  10  \\
Normal    &   0.35     &  0.5 \\
\hline
\end{tabular}
\end{center}
\end{table}

We used MATLAB optimization toolbox's \texttt{fmincon()} function for the implementation of the interior point algorithm (\cite{MATLAB2017b}) to solve problem (\ref{mple}). 
As in \cite{KoenkerMizera2014}, in Table \ref{table:EM-IP-time}  we report timing information and the values of $L(\bbeta)$ achieved for the interior point algorithm and EM algorithms 
with various number of iterations averaged over 100 replications. 

\begin{table}
\caption{Comparison of EM and interior point solutions: Iteration
counts, log likelihoods, and CPU times (in seconds) for three EM
variants and the interior point solver averaged over 100 replications}
\label{table:EM-IP-time}
\begin{center}
\begin{tabular} {l c c c c }
\hline
\hline
Algorithm & EM1 & EM2 & EM3 &  IP  \\
\hline
Iterations & 50 & 100 & 500 &  18  \\
Time       & 0.82 & 1.56 & 7.87 & 0.35 \\
$L(\bbeta)$-506 & 0.7880 & 1.0292 & 1.0780 & 5.7928 \\
\hline
\end{tabular}
\end{center}
\end{table}

The EM algorithm makes little progress from 50 to 500 iterations. 
By contrast, the interior point algorithm as implemented in MATLAB is both quicker and more
accurate. 

Table \ref{table:EM-IP-weight} reports initial values and root mean squared error (RMSE) of estimated component proportion parameters
by the algorithms,   
where the initial values for proportions are all equal.
The interior point algorithm's fitted proportions are closer to the true proportions than the ones by the EM algorithms.

\begin{table}
\begin{center}
\caption{Comparison of EM and interior point solutions: RMSE of proportion estimates}
\label{table:EM-IP-weight}
\begin{tabular} {l|l|l|l|l|l|l}
\hline
\hline
\multirow{2}{*}{Component}& True & Initial & 
     \multicolumn{4}{c}{RMSE of  Proportion Estimates} \\
        \cline{4-7}
          &   Proportion    & Value  &   EM1 & EM2 & EM3 & IP \\         
\hline
Clayton   &   0.40     &  0.33   &   0.0605 &  0.0604  &  0.0604  &  0.0195  \\
Gumbel    &   0.25     &  0.33   &   0.0775 & 0.0776 & 0.0776 & 0.0310   \\
Normal  &   0.35     &  0.33   &   0.0169 & 0.0171 & 0.0172 & 0.0115  \\
\hline
\end{tabular}
\end{center}
\end{table}

Table \ref{table:EM-IP-param}  reports 
RMSE of the component copula parameter estimates by the algorithms. 
The initial value of a copula parameter is its maximum likelihood estimate for the single component model without mixtures.
The interior point algorithm's fitted copula parameters are closer to the true copula parameters than the ones by the EM algorithms. 

\begin{table}
\begin{center}
\caption{Comparison of EM and interior point solutions: RMSE of copula parameter estimates}
\label{table:EM-IP-param}
\begin{tabular} {l|c|c|c|c|c}
\hline
\hline
\multirow{2}{*}{Component}& True &  
     \multicolumn{4}{c}{RMSE of  copula parameter estimates} \\
        \cline{3-6}
          &  Parameter   &   EM1 & EM2 & EM3 & IP \\         
\hline
Clayton   &    3       & 0.2586 & 0.2206 & 0.2135 & 0.1490  \\
Gumbel    &   10       & 4.3368 & 4.2931 & 4.2854 & 0.1372 \\
Normal  &   0.5      &0.0084 & 0.0098 & 0.0101 & 0.0073 \\
\hline
\end{tabular}
\end{center}
\end{table}

\section{Monte Carlo Simulations} \label{sec:simu}

We conduct Monte Carlo simulations to examine the finite sample performance
of the proposed CFGTN estimator and to compare it with a kernel copula density estimator.
The kernel copula density estimator is implemented in the R package \texttt{np} (\cite{HayfieldRacine_JSS_Jul2008}) using a 
normal kernel and bandwidth selected by the normal reference rule-of-thumb.

Because our model selection strategy starts with the best single component parametric model among the Clayton, Frank, Gumbel, T, or normal copula according to the  model selection criterion, the estimated copula density achieves best possible outcome if the true copula is from one of these 5 families. Therefore we omit this scenario in the simulation.  

We include in our simulations two groups of copulas: group one for nested models;  group two for non-nested models.  Specifically, they are:

\emph{Group 1: nested models}
\begin{itemize}
	\item Clayton-Frank: mixture of Clayton and Frank copulas;
	\item Clayton-$\text{T}_5$: mixture of Clayton and T with 5 degrees of freedom (DoF) copulas;  
	\item Clayton-Normal: mixture of Clayton and normal copulas;
	\item Clayton-Frank-Gumbel-$\text{T}_5$-Normal: mixture of Clayton, Frank, Gumbel,  T with 5 DoF, and normal copulas.
\end{itemize}

\emph{Group 2: non-nested models}
\begin{itemize}
	\item Clayton-$\text{T}_5$-$\text{T}_{15}$: mixture of Clayton, T with 5 DoF, and T with 15 DoF copulas;
	\item $\text{T}_5$-$\text{T}_{15}$: mixture of T with 5 DoF and T with 15 DoF copulas;  
	\item $\text{T}_5$-$\text{T}_{15}$-Normal: mixture of T with 5 DoF, T with 15 DoF, and normal copulas. 
\end{itemize}

For each copula distribution, we consider four levels of dependence with
Kendall's $\tau$ being 0.2, 0.4, 0.6 and 0.8 for each copula component, respectively. We assign equal proportion for each component of the mixture model. For example, for the case of Clayton-Frank model with Kendall's $\tau$ being 0.2, (1) the mixture proportion for the Clayton component is $0.5$ and the Clayton copula parameter is determined by the requirement that the Kendall's $\tau$ for the Clayton copula is 0.2; (2) the mixture proportion for the Frank component is $0.5$ and the Frank copula parameter is determined by the requirement that the Kendall's $\tau$ for the Frank copula is 0.2.  

\begin{figure}[!ht]
	\centering
	\scalebox{0.5}{\includegraphics{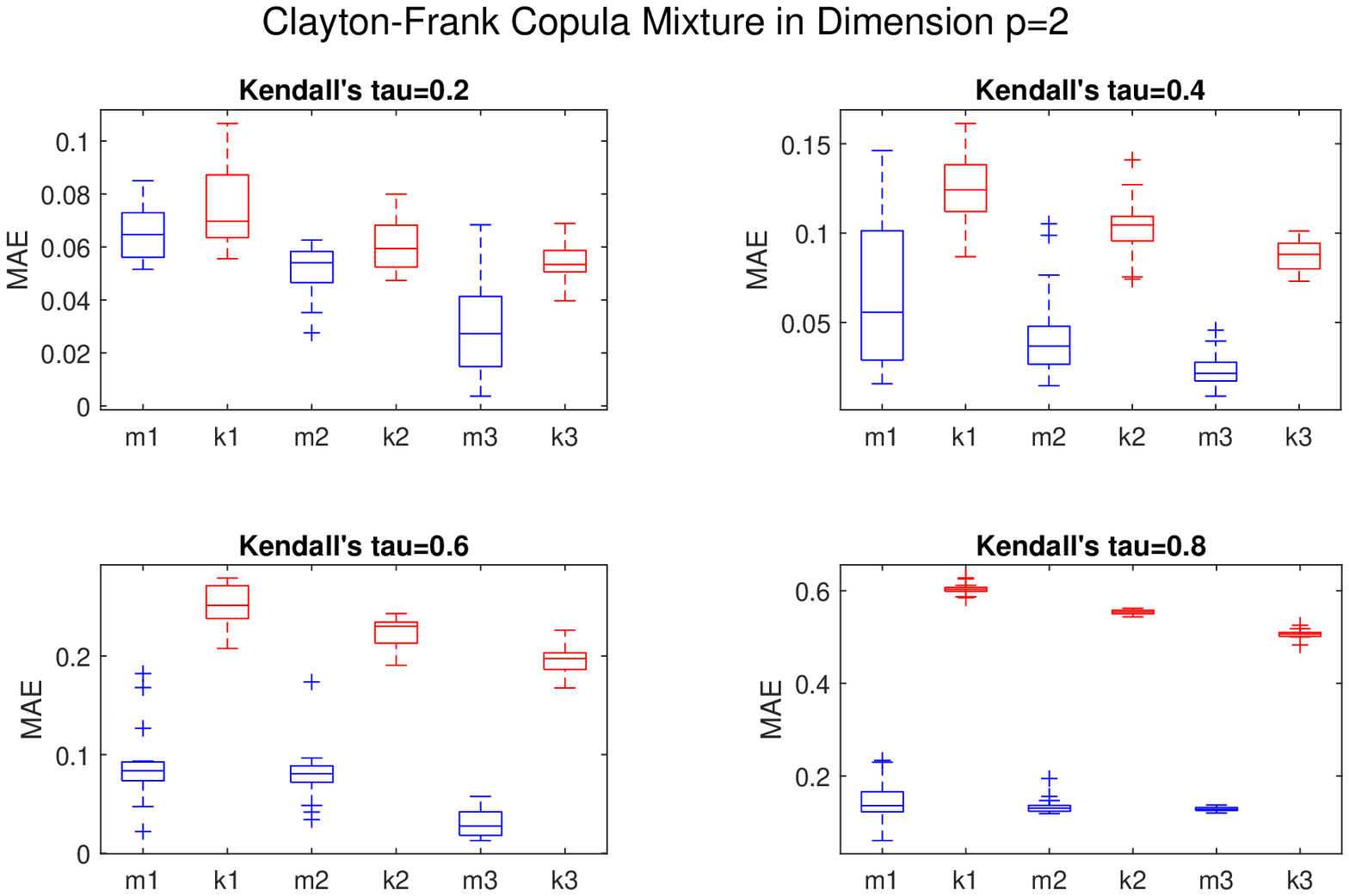}}
	\scalebox{0.5}{\includegraphics{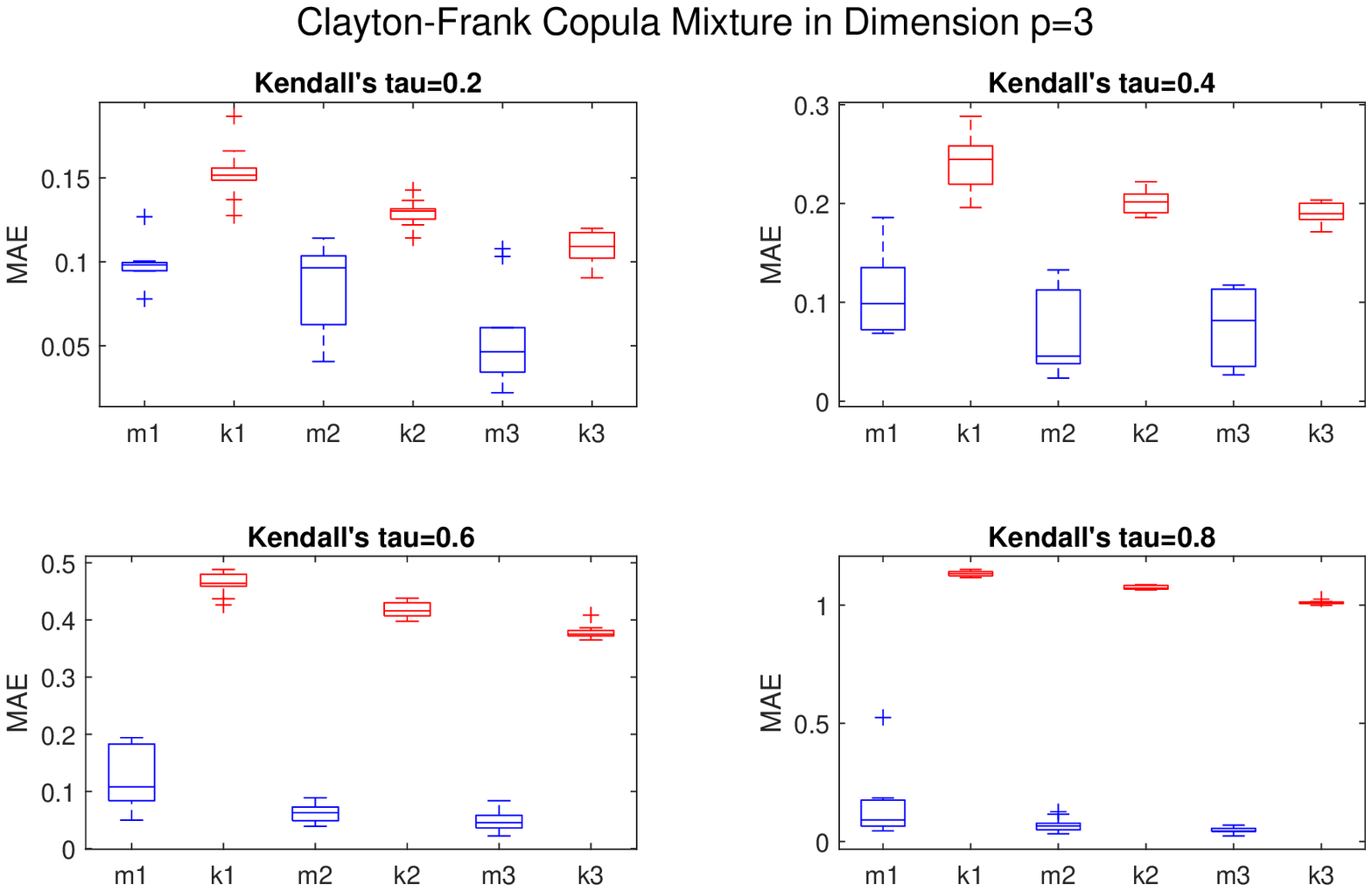}}
	\scalebox{0.5}{\includegraphics{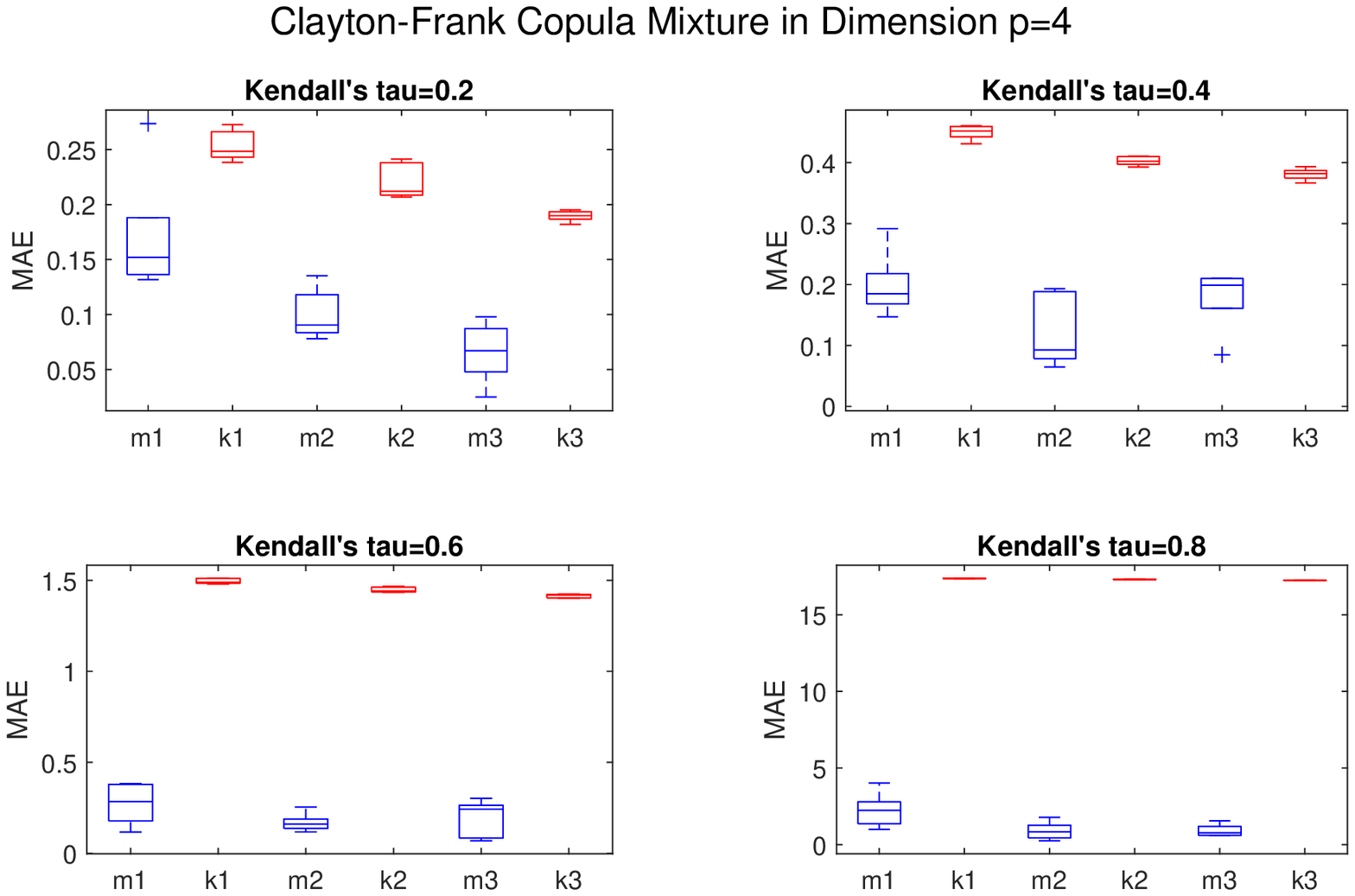}}
	\caption{Boxplots of the mean absolute error (MAE)  for the Clayton-Frank mixture copula in dimension $p=2, 3, 4$ by the \textcolor{blue}{CFGTN} copula density estimator 
		as indicated by the letter `m' in the x-axis label and \textcolor{red}{kernel} copula density estimator as indicated by the letter `k' for sample sizes $n=500, 1000, 2000$ as indicated by numbers 1, 2, 3 respectively in the x-axis label}
	\label{CF}
\end{figure}

\begin{figure}[!ht]
	\centering
	\scalebox{0.5}{\includegraphics{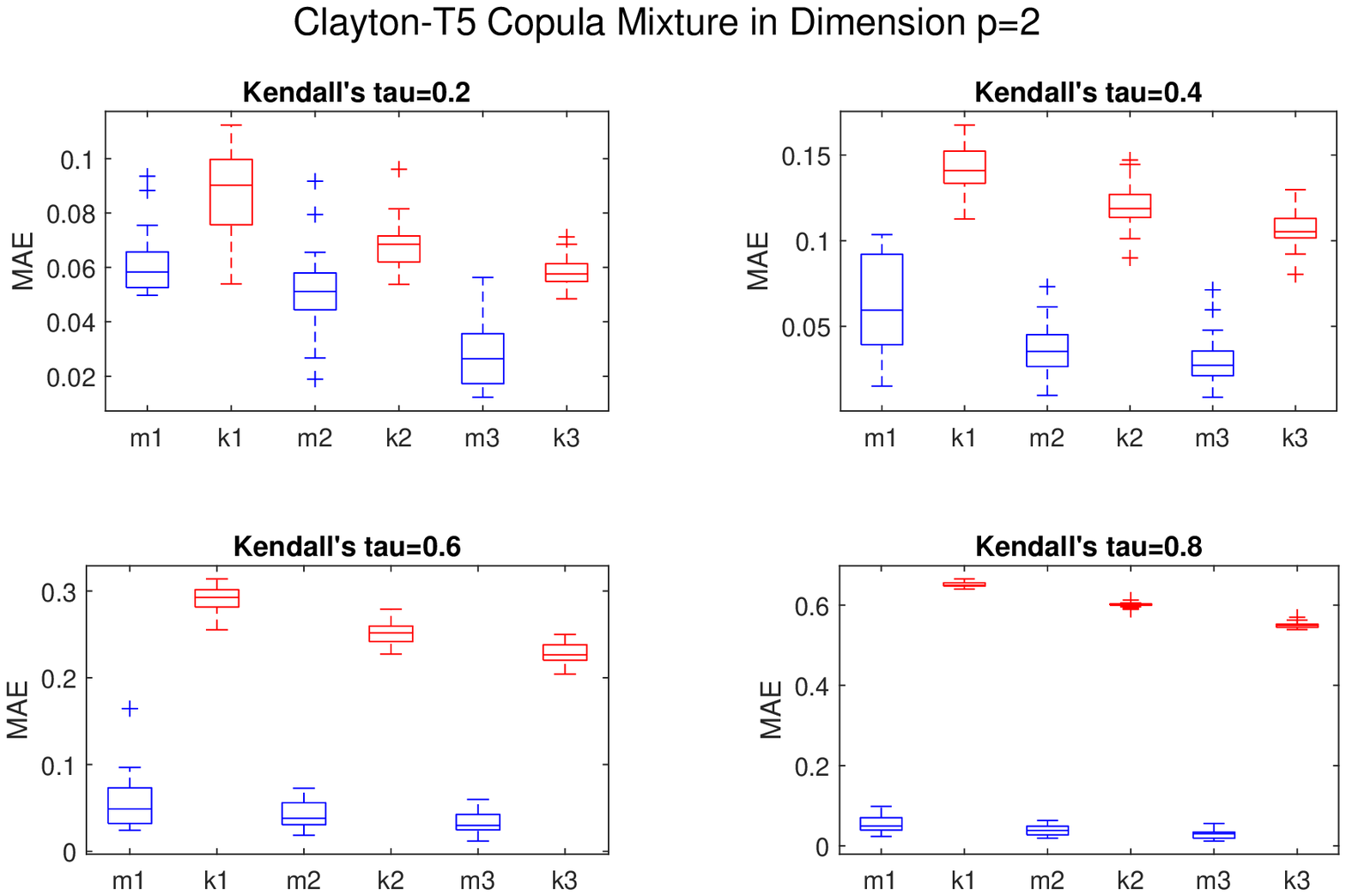}}
	\scalebox{0.5}{\includegraphics{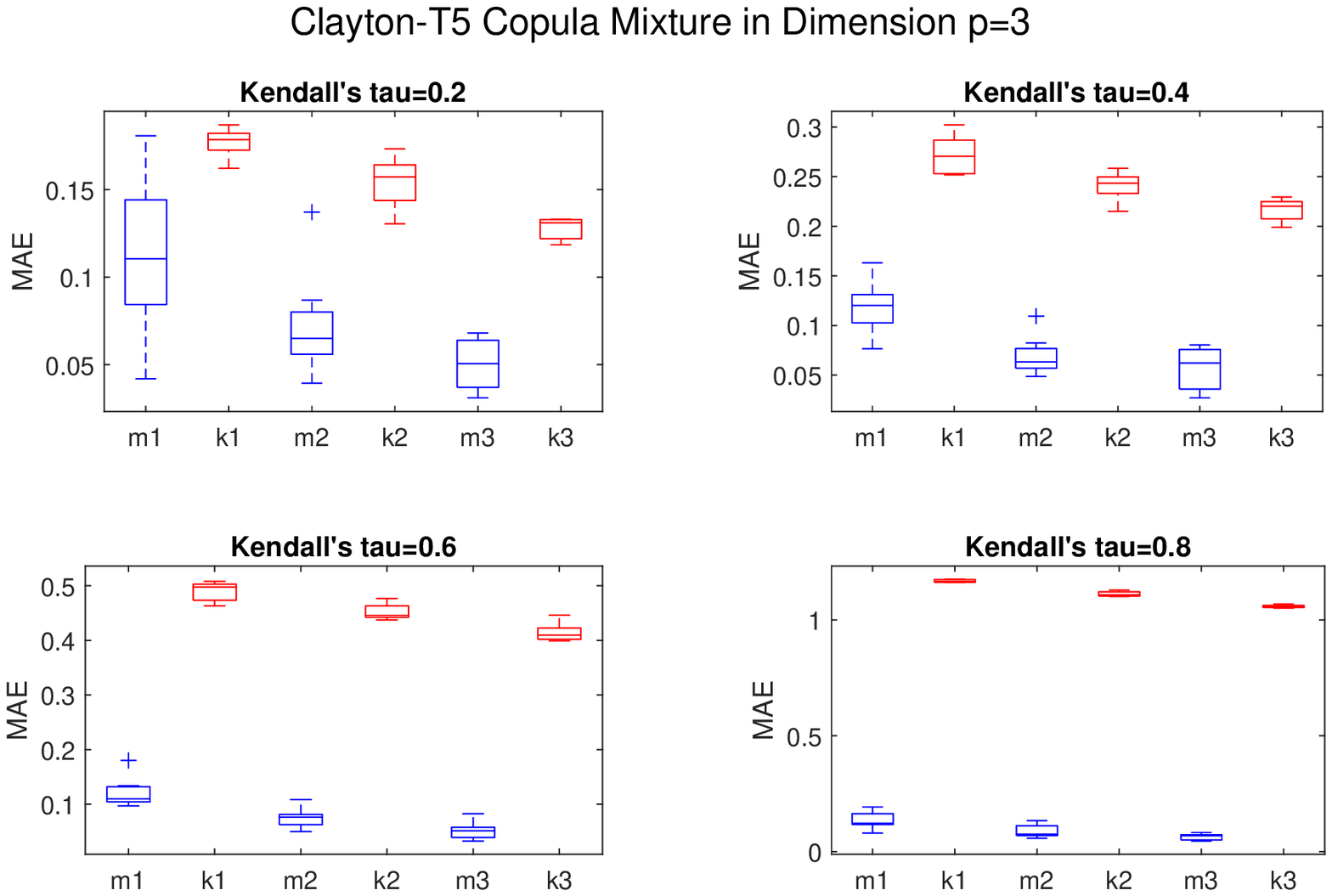}}
	\scalebox{0.5}{\includegraphics{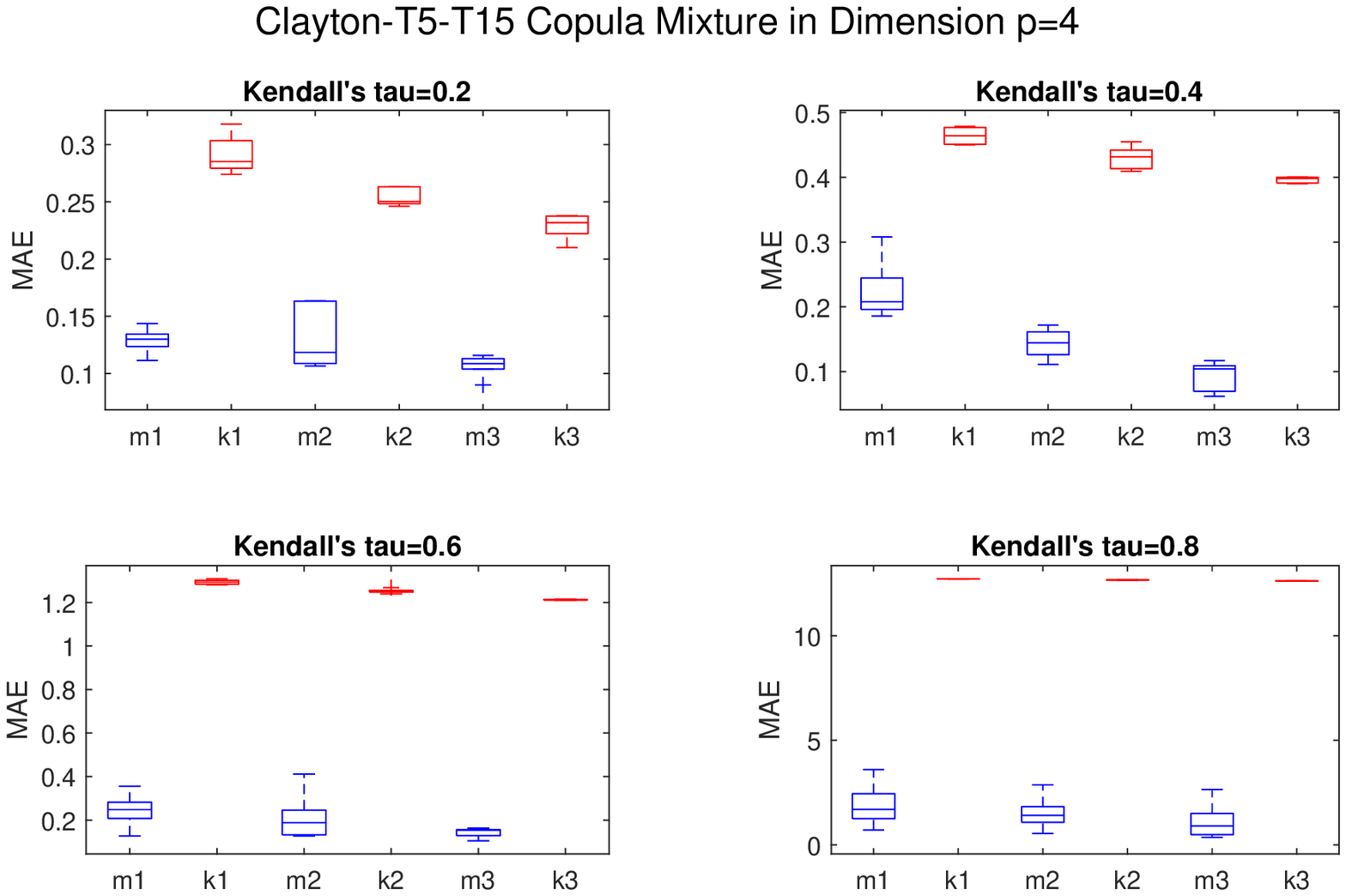}}
	\caption{Boxplots of the mean absolute error (MAE)  for the Clayton-$\text{T}_5$ mixture copula in dimension $p=2, 3, 4$ by the \textcolor{blue}{CFGTN} copula density estimator 
		as indicated by the letter `m' in the x-axis label and \textcolor{red}{kernel} copula density estimator  as indicated by the letter `k' for sample sizes $n=500, 1000, 2000$ as indicated by numbers 1, 2, 3 respectively in the x-axis label}
	\label{CT5}
\end{figure}

\begin{figure}[!ht]
	\centering
	\scalebox{0.5}{\includegraphics{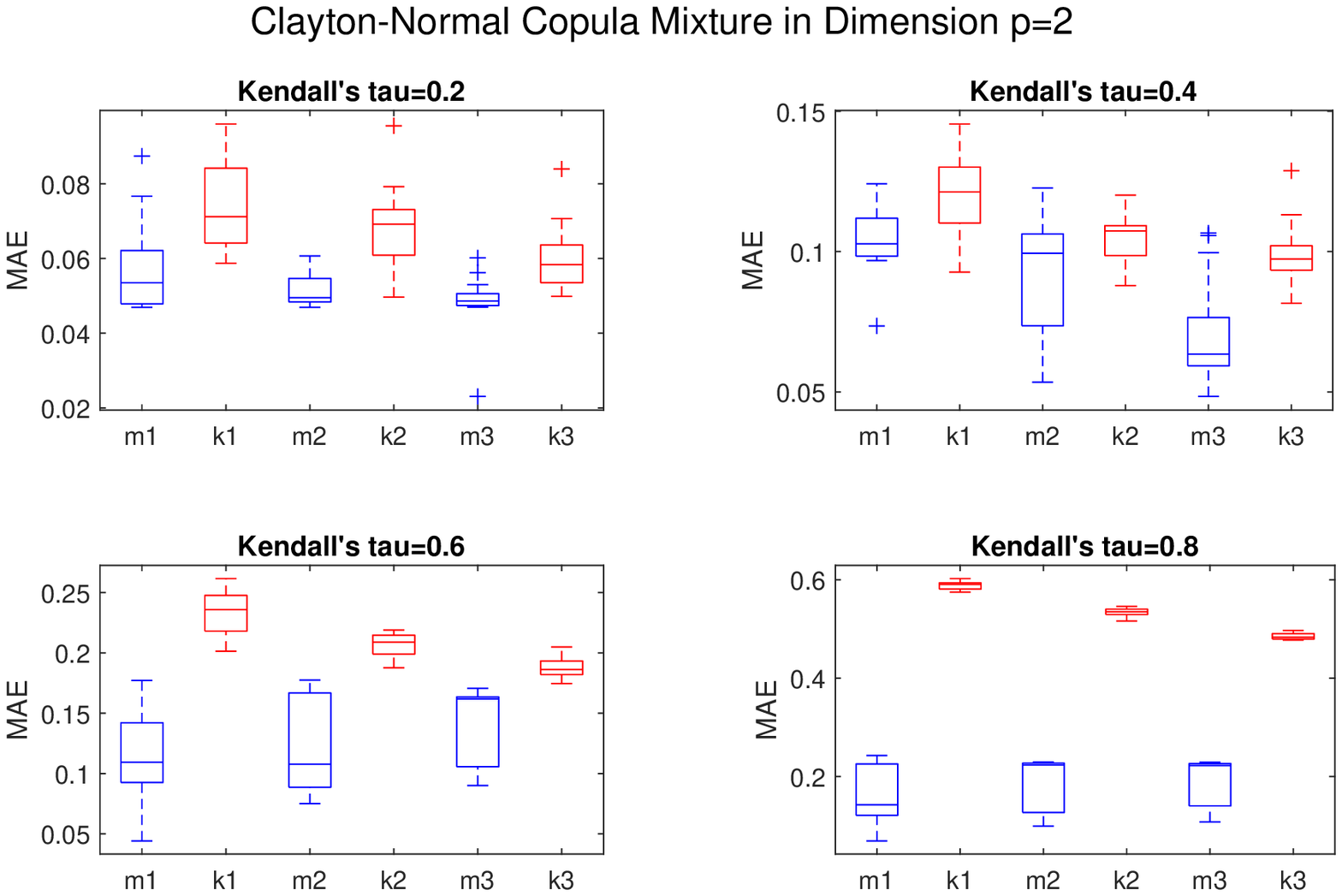}}
	\scalebox{0.5}{\includegraphics{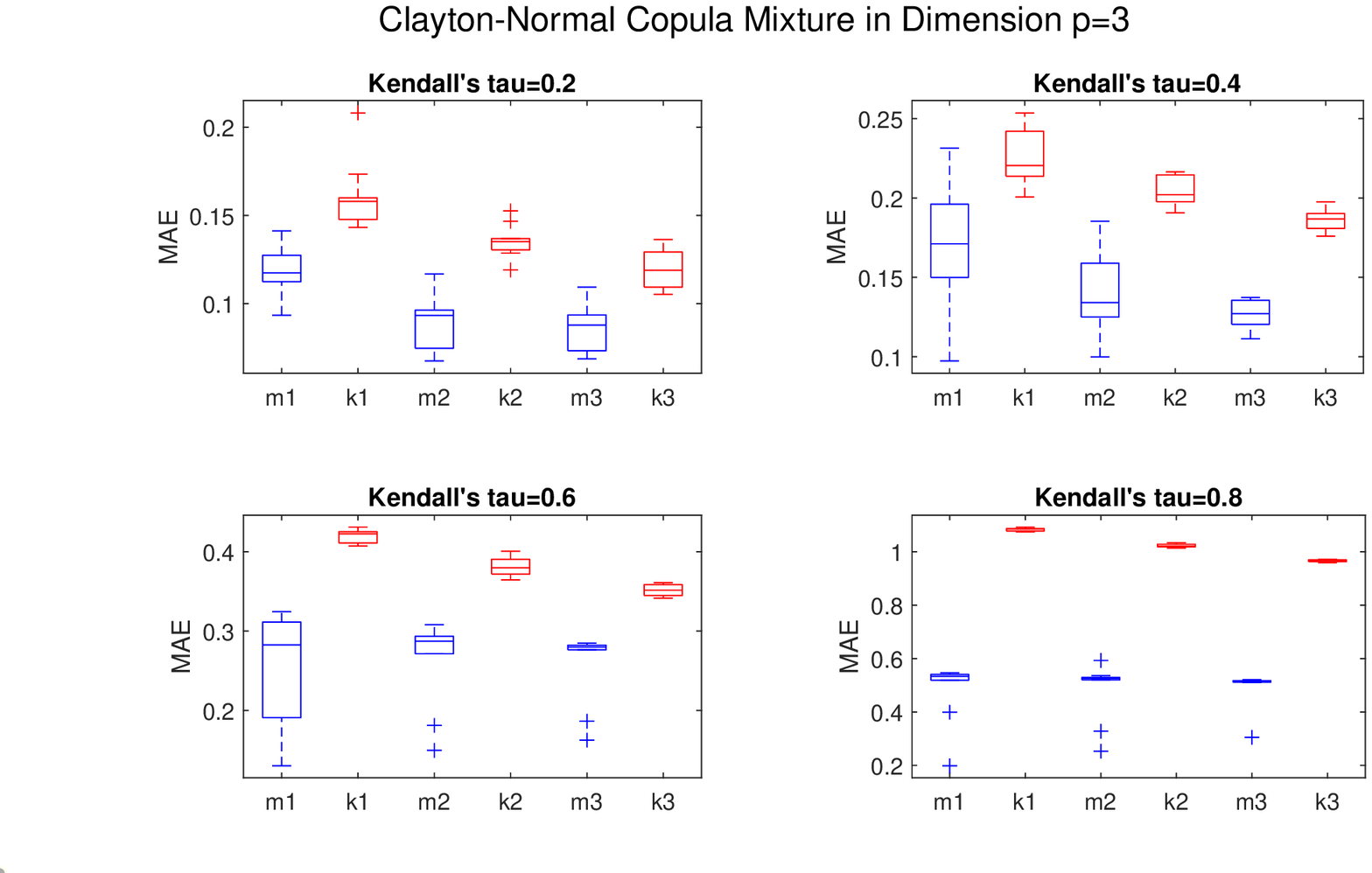}}
	\scalebox{0.5}{\includegraphics{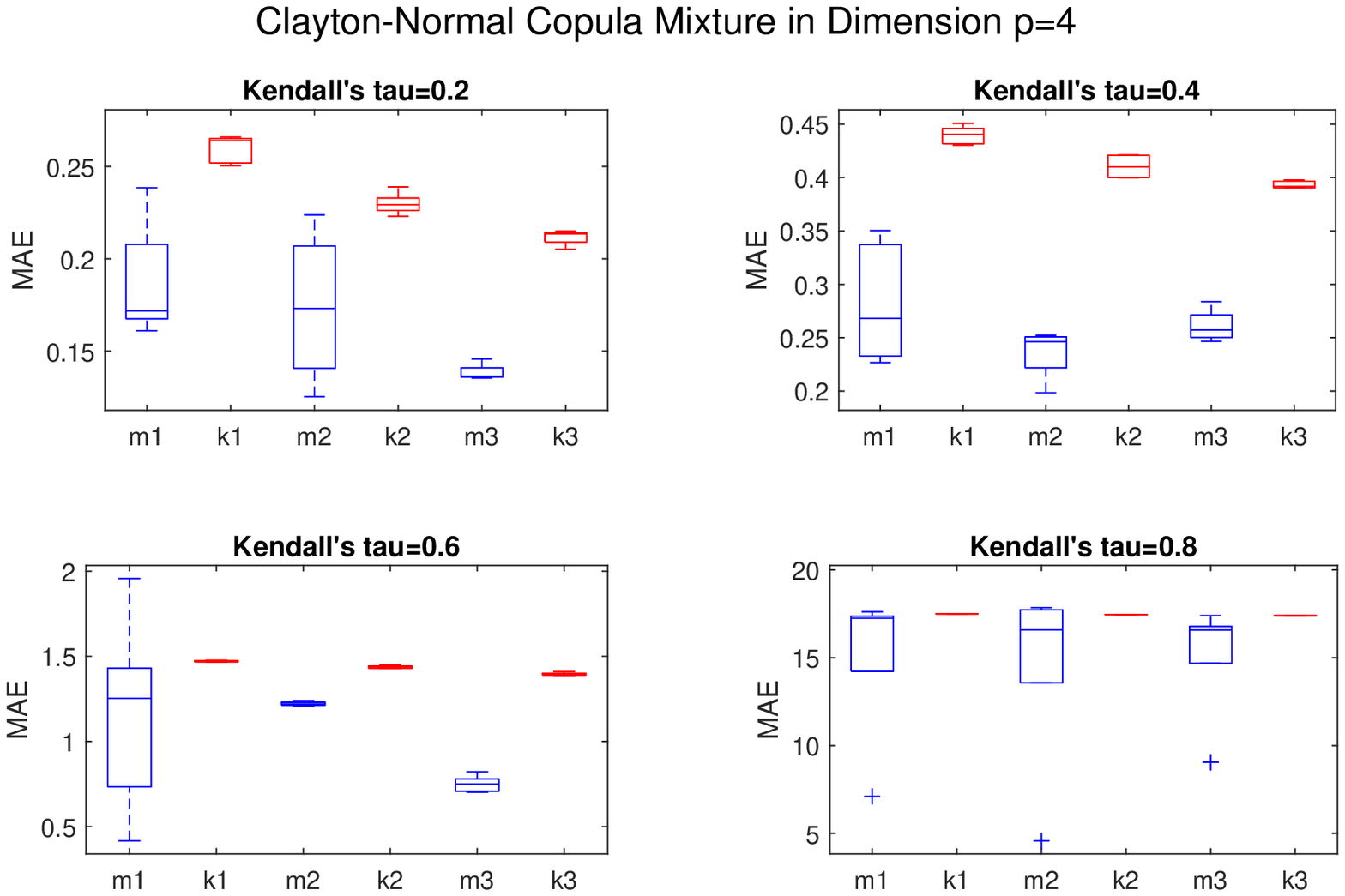}}
	\caption{Boxplots of the mean absolute error (MAE)  for the Clayton-Normal mixture copula in dimension $p=2, 3, 4$ by the \textcolor{blue}{CFGTN} copula density estimator 
		as indicated by the letter `m' in the x-axis label and \textcolor{red}{kernel} copula density estimator  as indicated by the letter `k' for sample sizes $n=500, 1000, 2000$ as indicated by numbers 1, 2, 3 respectively in the x-axis label}
	\label{CN}
\end{figure}

\begin{figure}[!ht]
	\centering
	\scalebox{0.5}{\includegraphics{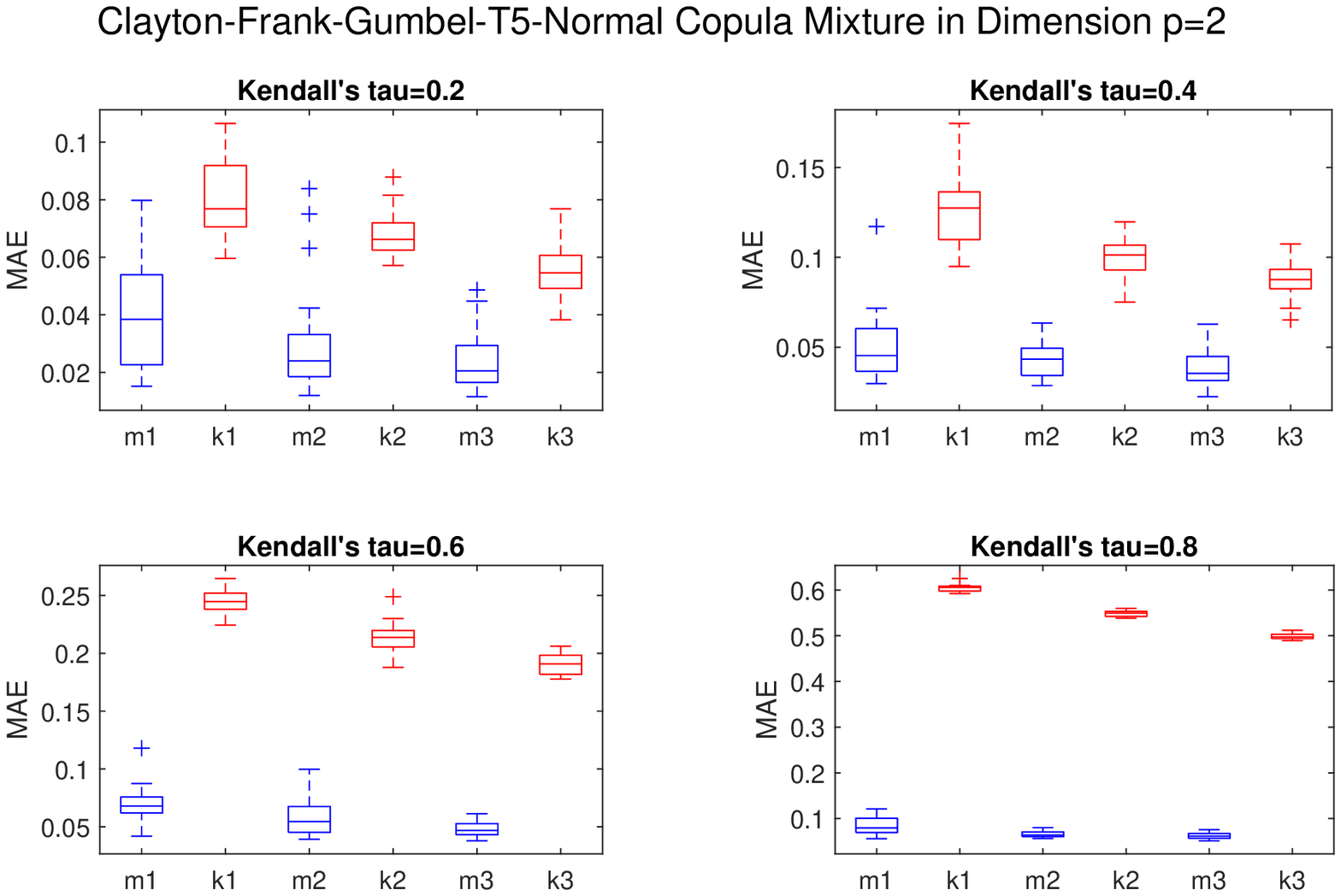}}
	\scalebox{0.5}{\includegraphics{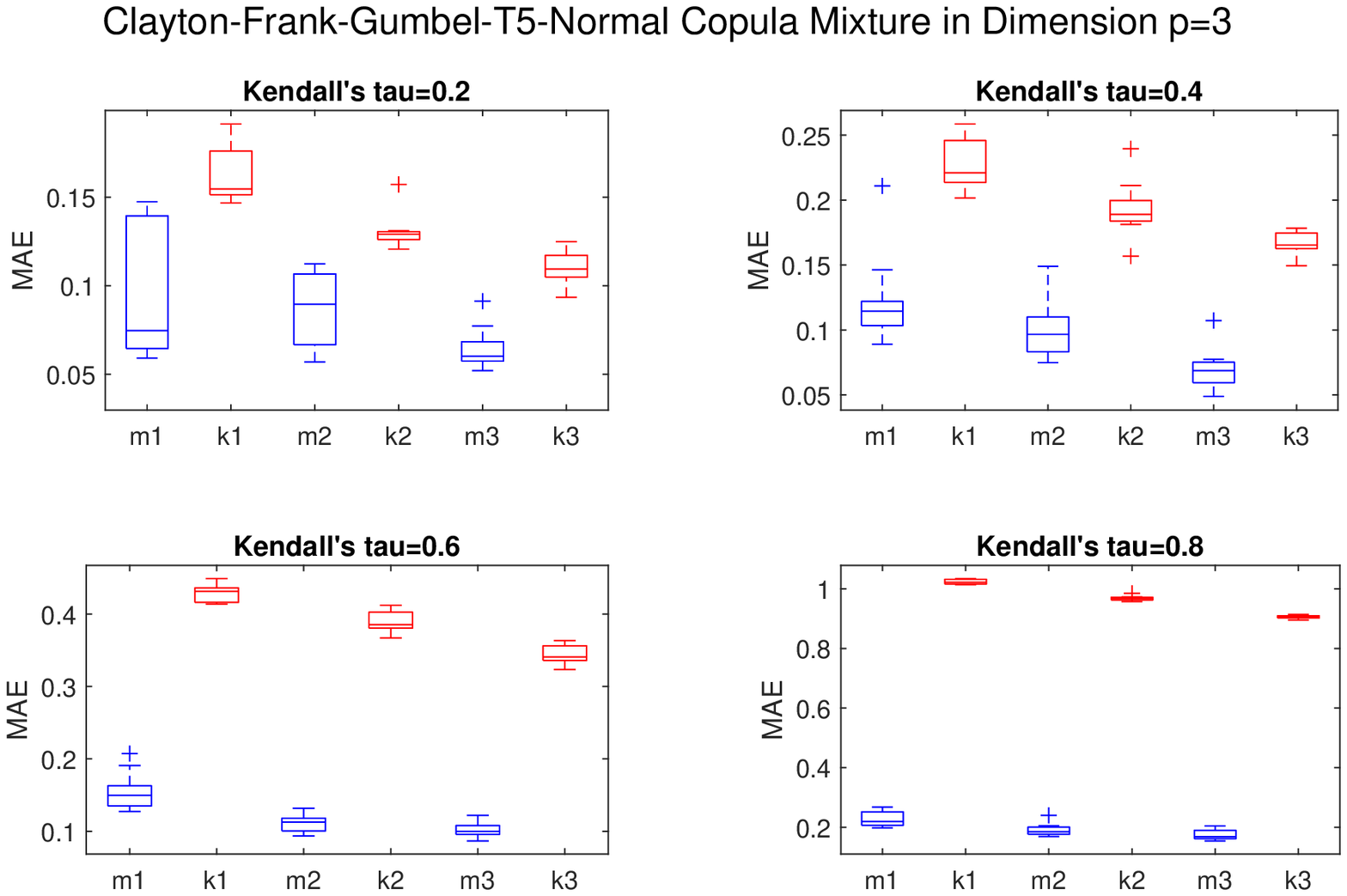}}
	\scalebox{0.5}{\includegraphics{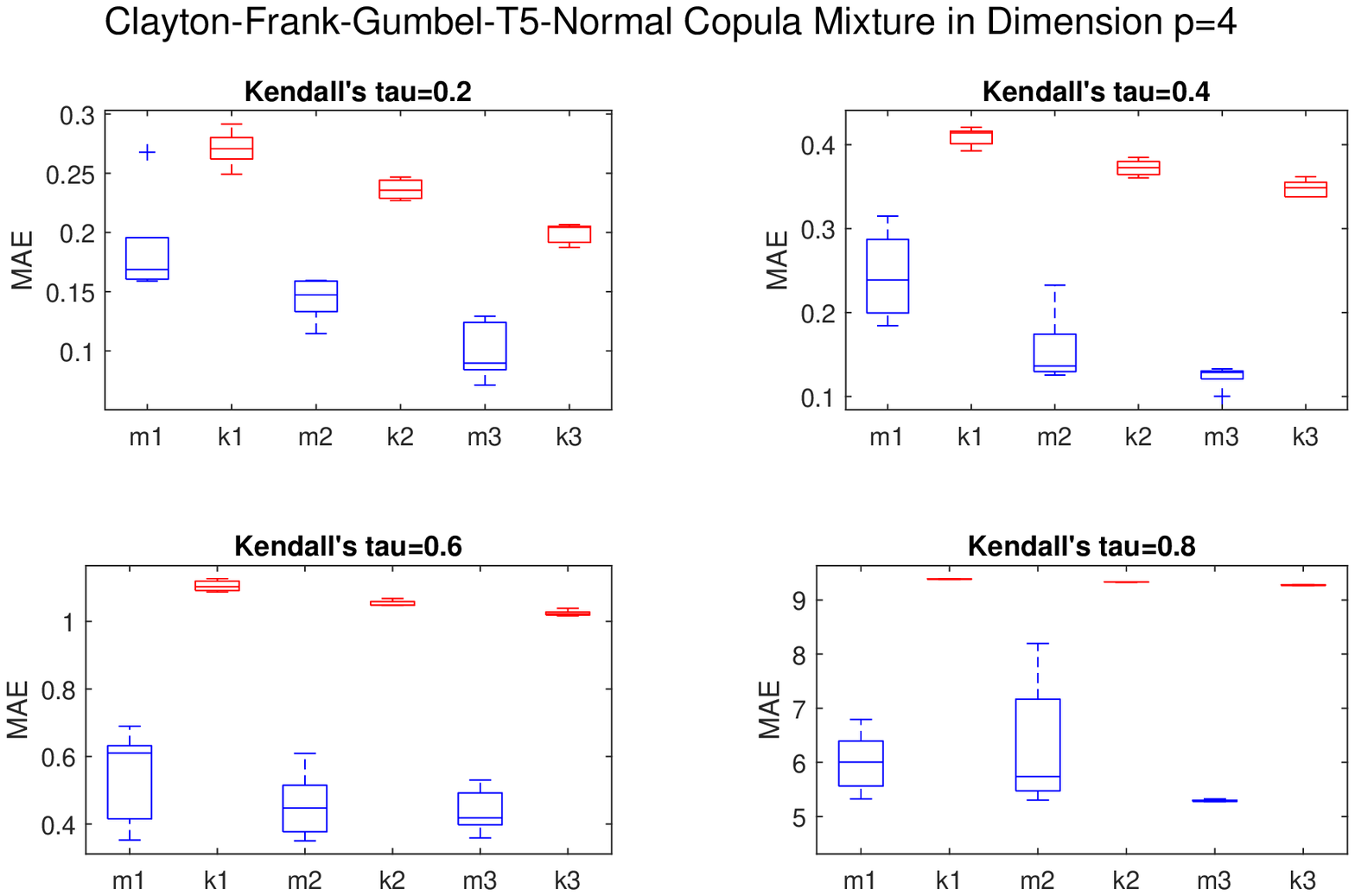}}
	\caption{Boxplots of the mean absolute error (MAE)  for the Clayton-Frank-Gumbel-$\text{T}_5$-Normal mixture copula in dimension $p=2, 3, 4$ by the \textcolor{blue}{CFGTN} copula density estimator as indicated by the letter `m' in the x-axis label  and \textcolor{red}{kernel} copula density estimator  as indicated by the letter `k' for sample sizes $n=500, 1000, 2000$ as indicated by numbers 1, 2, 3 respectively in the x-axis label}
	\label{CFGT5N}
\end{figure}

\begin{figure}[!ht]
	\centering
	\scalebox{0.5}{\includegraphics{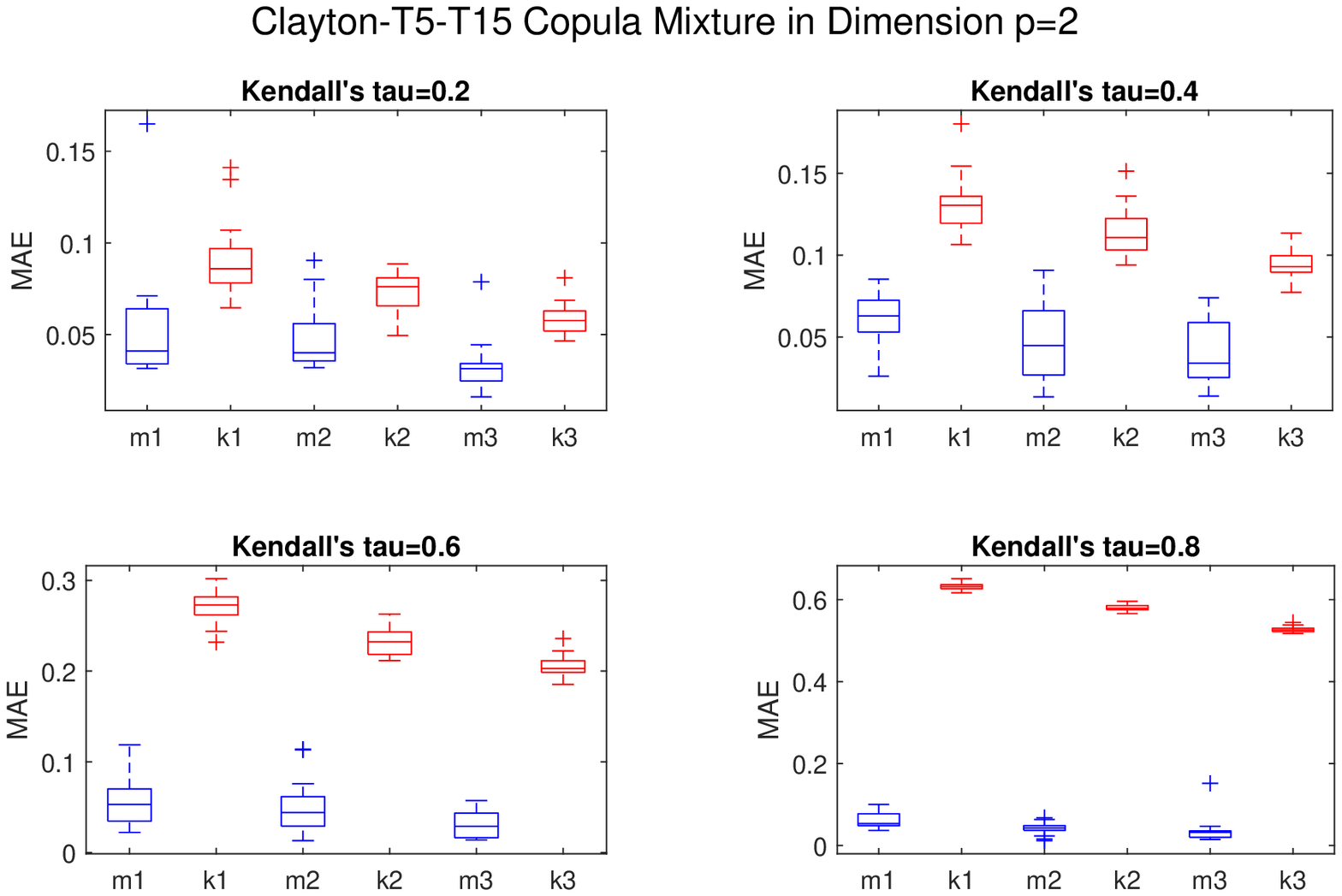}}
	\scalebox{0.5}{\includegraphics{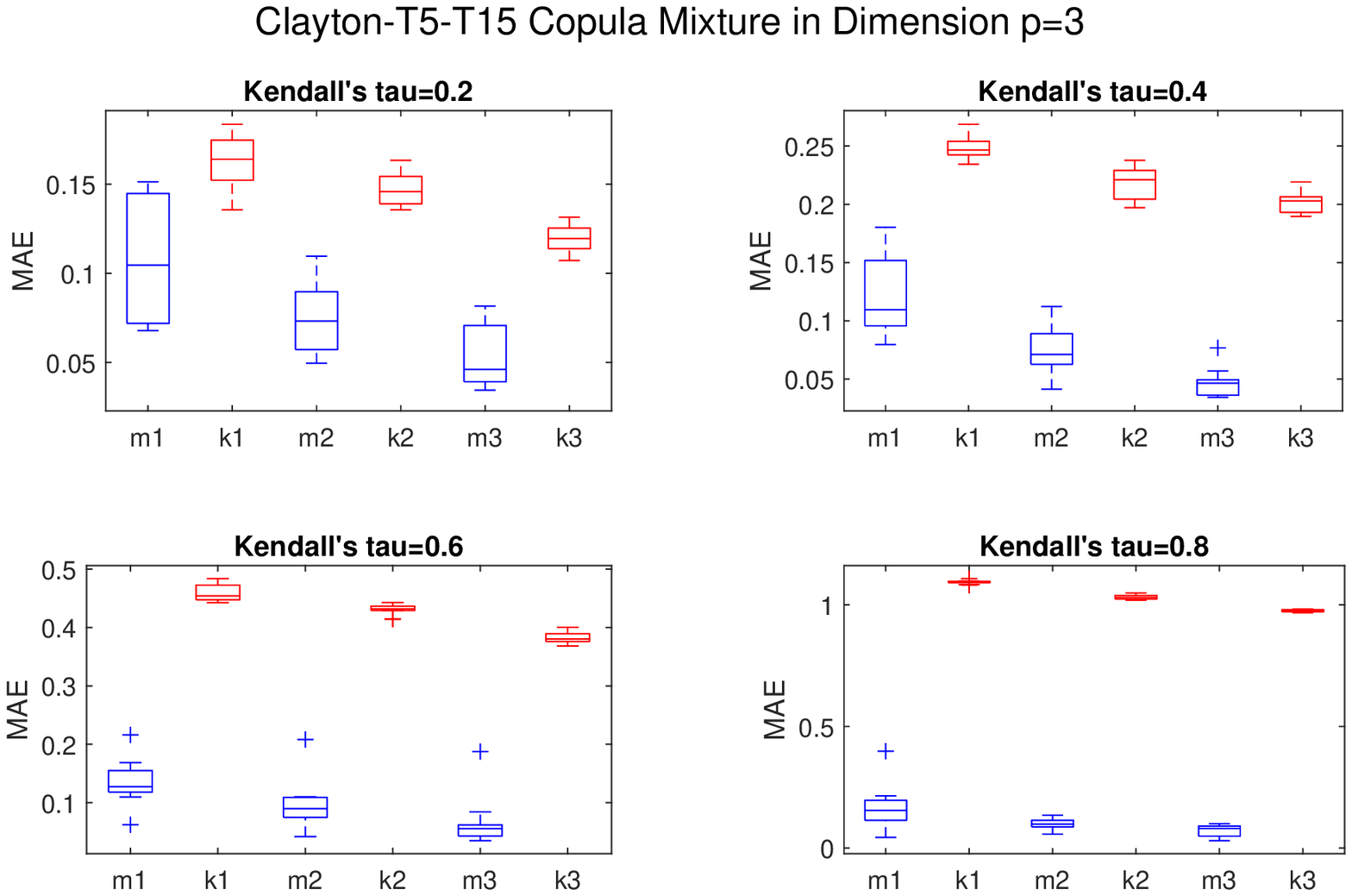}}
	\scalebox{0.5}{\includegraphics{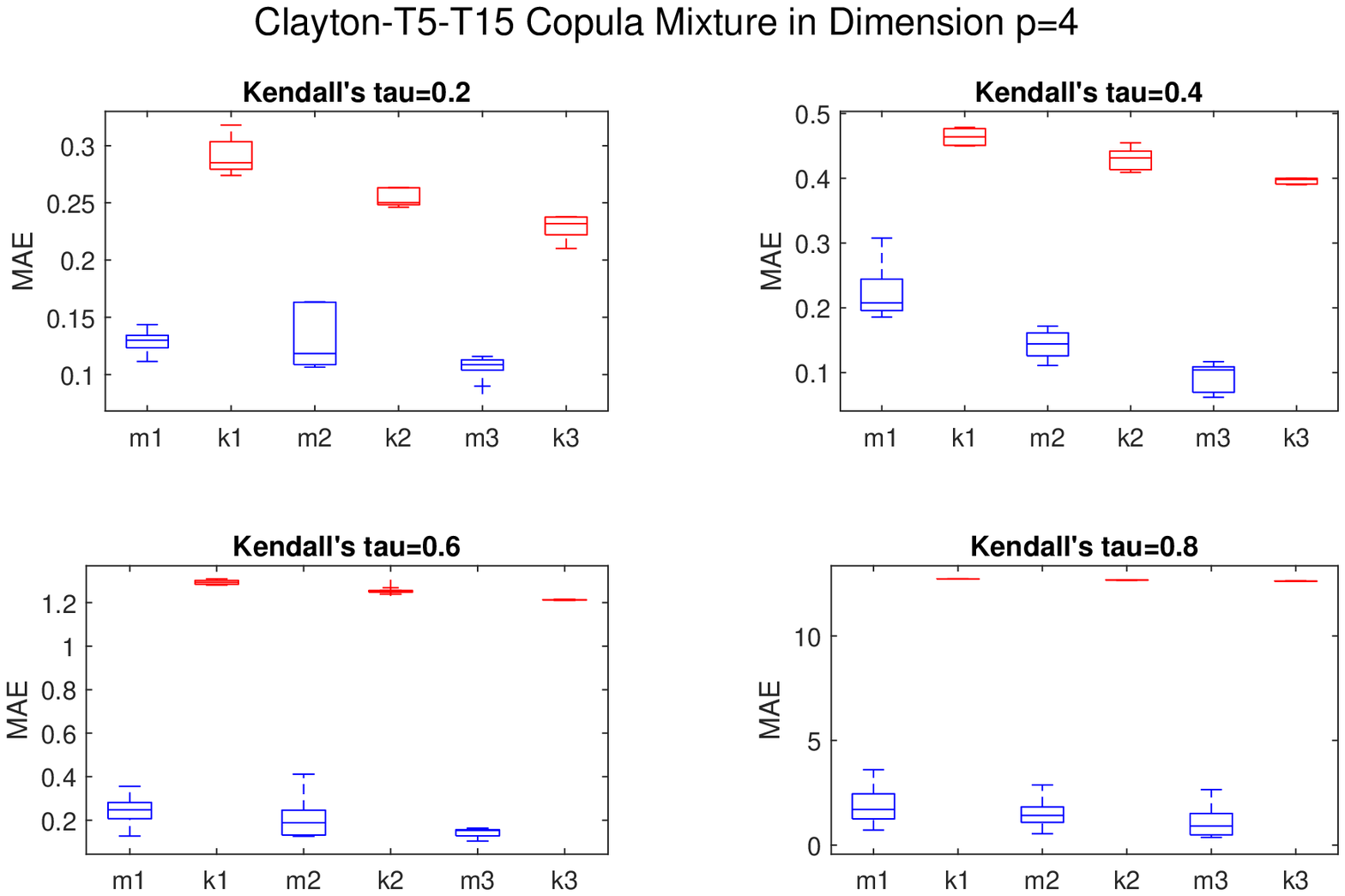}}
	\caption{Boxplots of the mean absolute error (MAE)  for the Clayton-$\text{T}_5$-$\text{T}_{15}$ mixture copula in dimension $p=2, 3, 4$ by the \textcolor{blue}{CFGTN} copula density estimator as indicated by the letter `m' in the x-axis label and \textcolor{red}{kernel} copula density estimator  as indicated by the letter `k' for sample sizes $n=500, 1000, 2000$ as indicated by numbers 1, 2, 3 respectively in the x-axis label}
	\label{CT5T15}
\end{figure}

\begin{figure}[!ht]
	\centering
	\scalebox{0.5}{\includegraphics{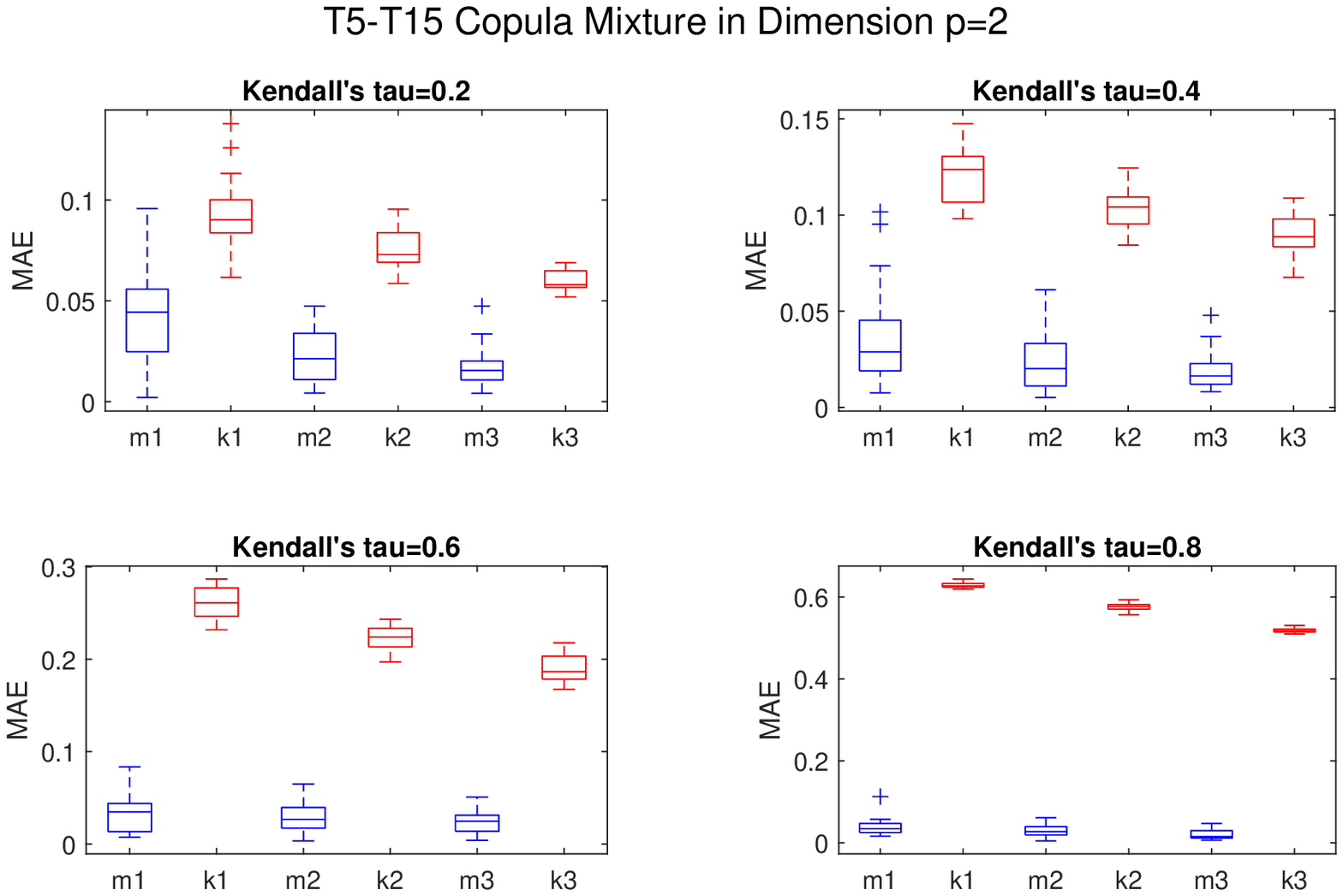}}
	\scalebox{0.5}{\includegraphics{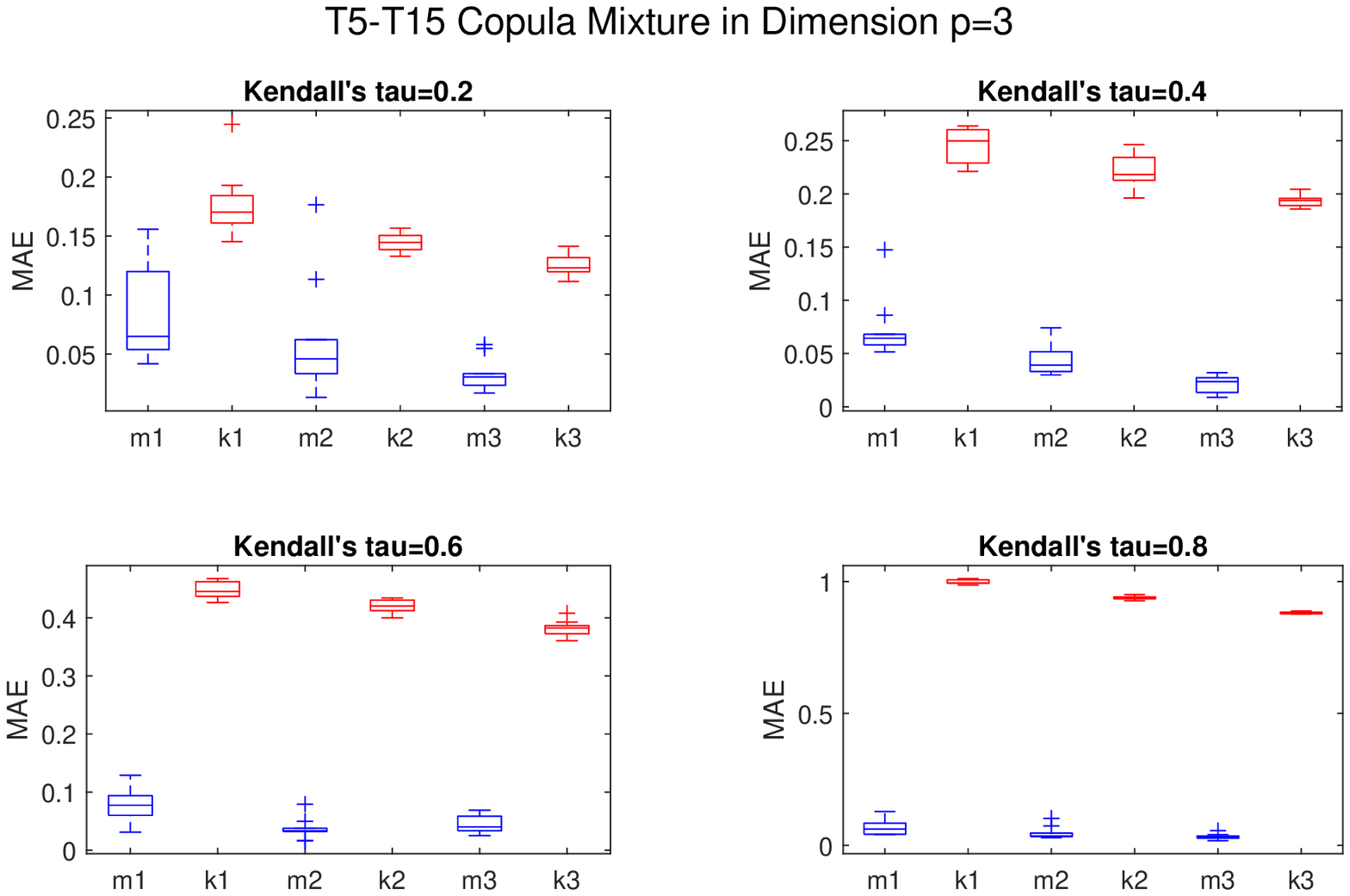}}
	\scalebox{0.5}{\includegraphics{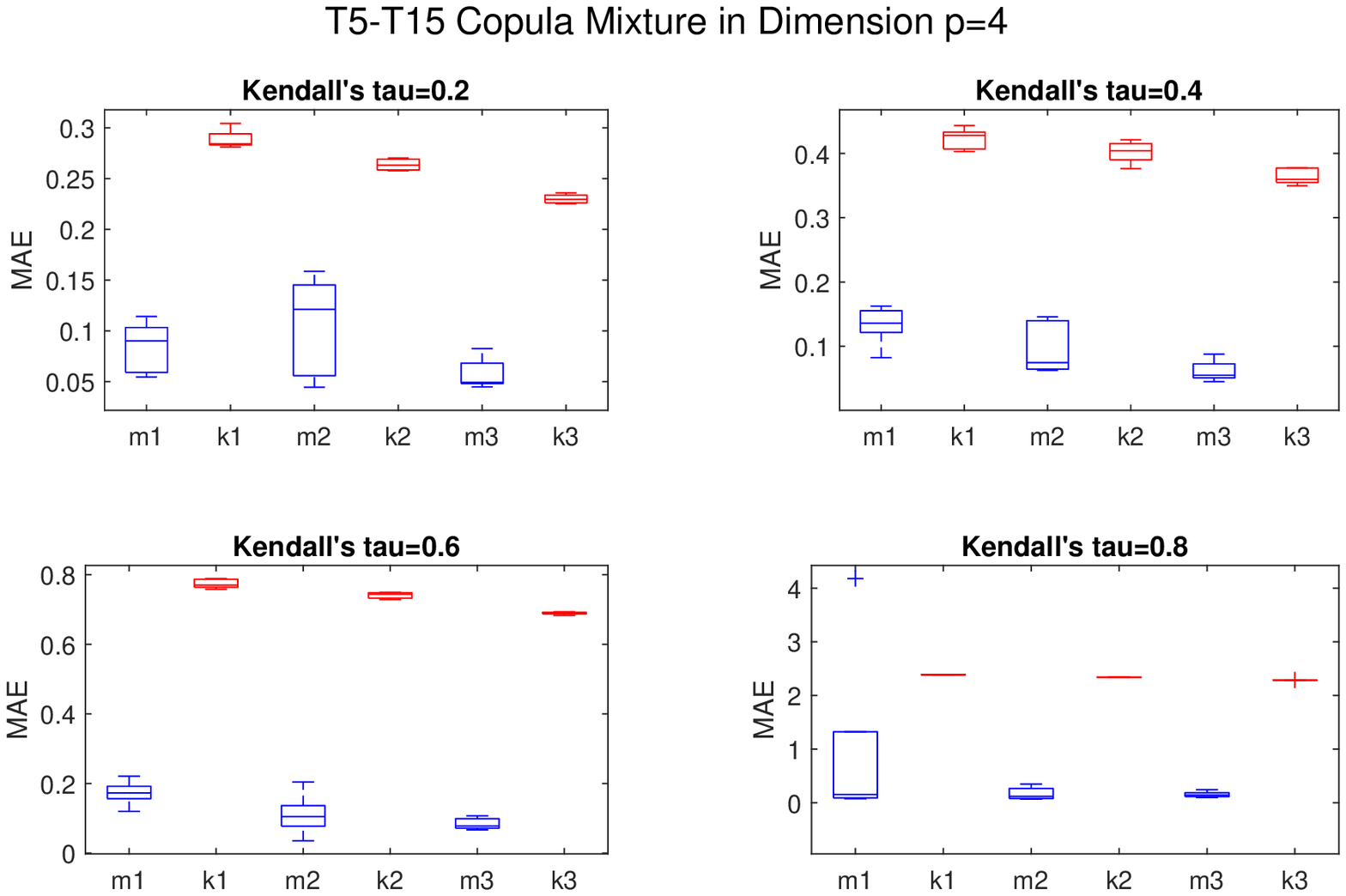}}
	\caption{Boxplots of the mean absolute error (MAE)  for the $\text{T}_5$-$\text{T}_{15}$ mixture copula in dimension $p=2, 3, 4$ by the \textcolor{blue}{CFGTN} copula density estimator as indicated by the letter `m' in the x-axis label and \textcolor{red}{kernel} copula density estimator  as indicated by the letter `k' for sample sizes $n=500, 1000, 2000$ as indicated by numbers 1, 2, 3 respectively in the x-axis label}
	\label{T5T15}
\end{figure}

\begin{figure}[!ht]
	\centering
	\scalebox{0.5}{\includegraphics{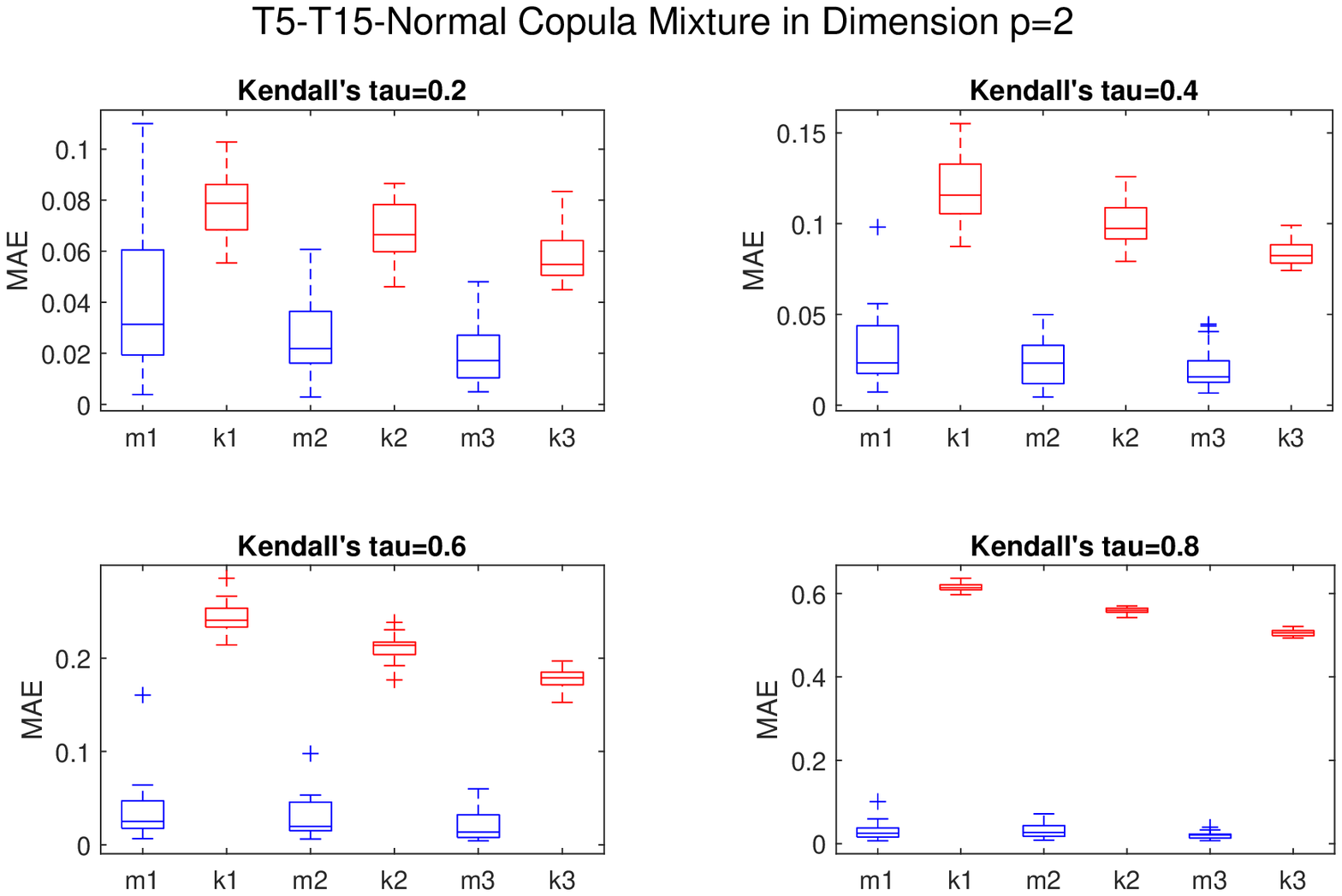}}
	\scalebox{0.5}{\includegraphics{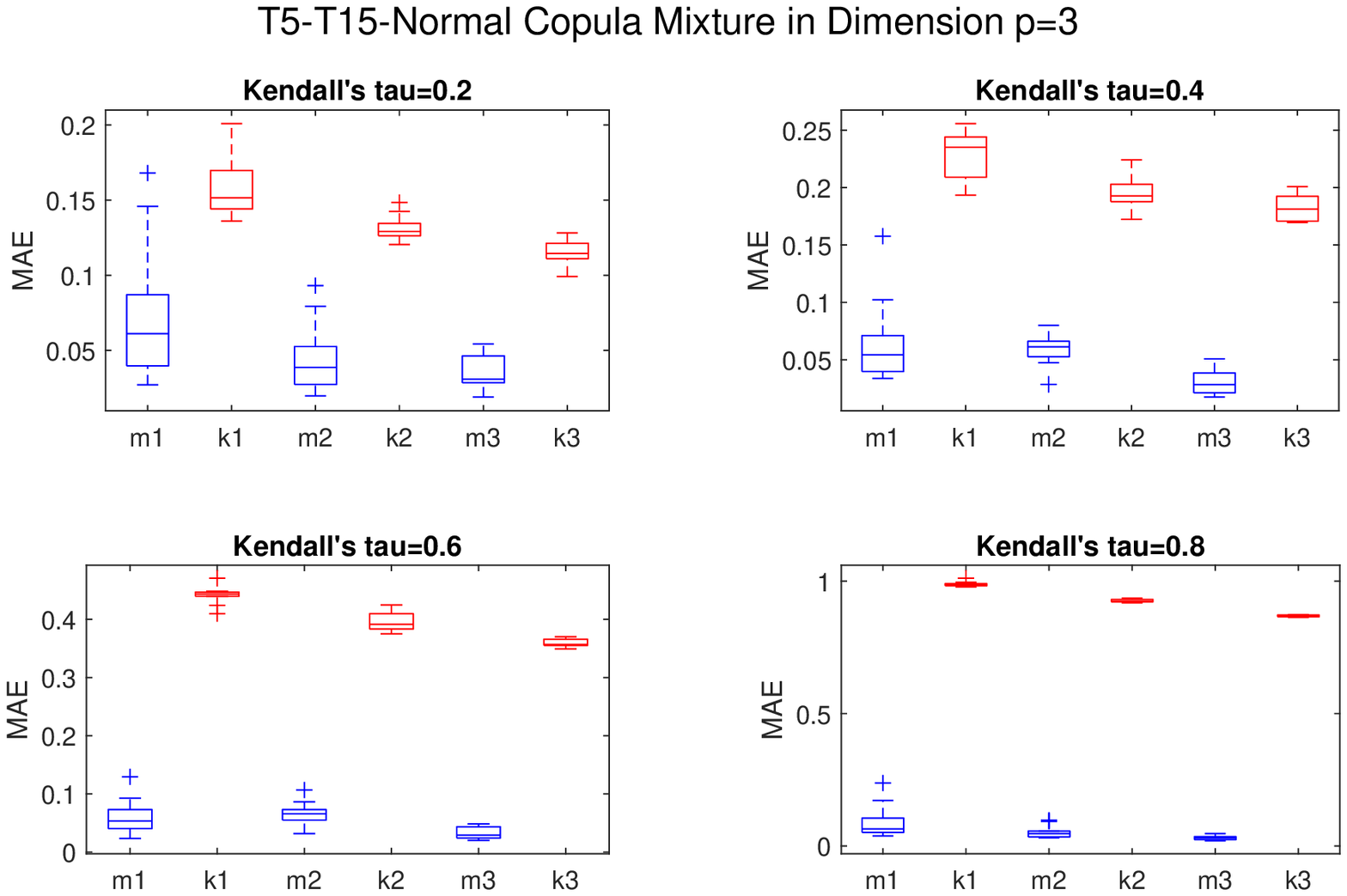}}
	\scalebox{0.5}{\includegraphics{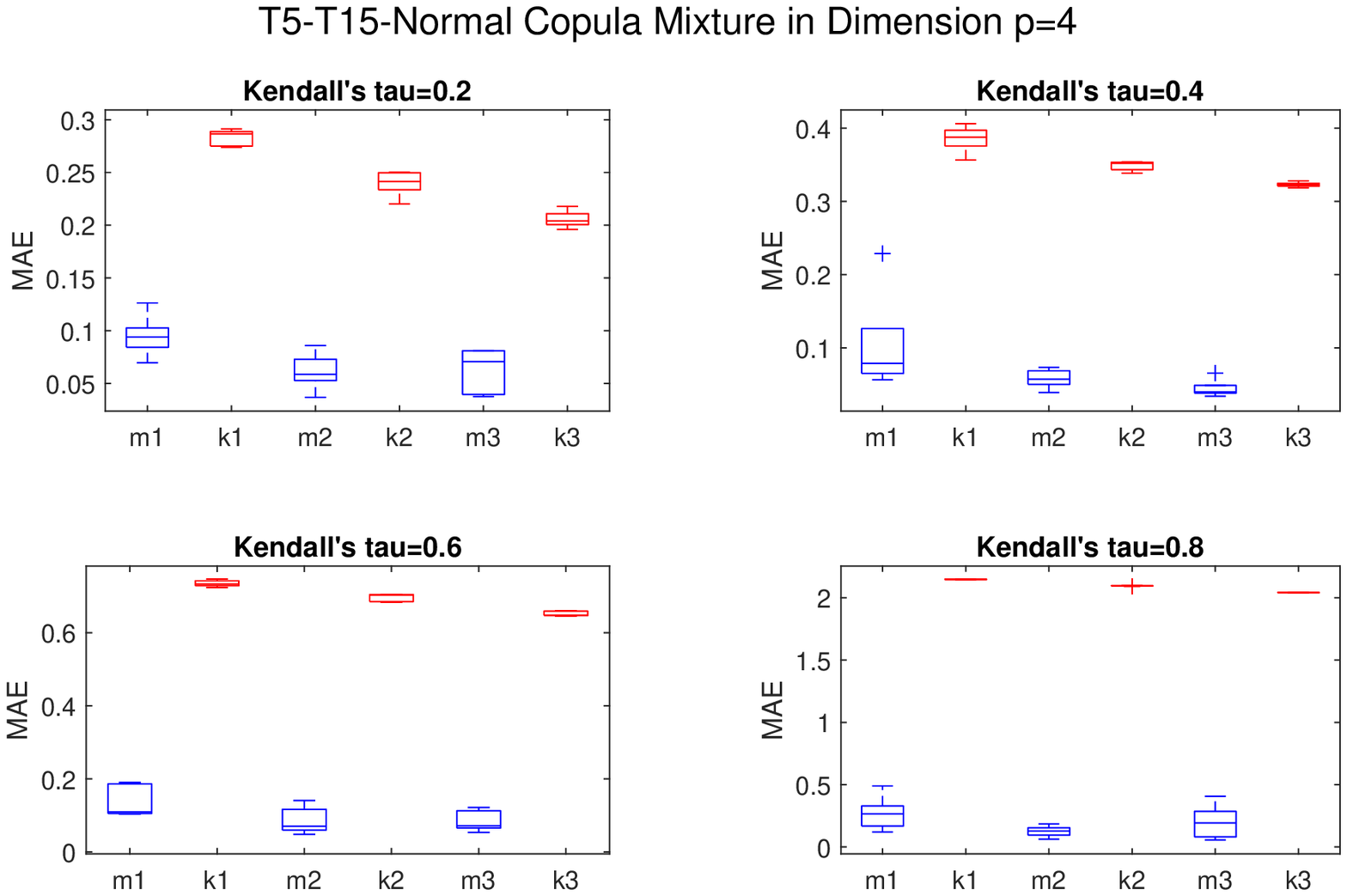}}
	\caption{Boxplots of the mean absolute error (MAE)  for the $\text{T}_5$-$\text{T}_{15}$-Normal mixture copula in dimension $p=2, 3, 4$ by the \textcolor{blue}{CFGTN} copula density estimator as indicated by the letter `m' in the x-axis label and \textcolor{red}{kernel} copula density estimator  as indicated by the letter `k' for sample sizes $n=500, 1000, 2000$ as indicated by numbers 1, 2, 3 respectively in the x-axis label}
	\label{T5T15N}
\end{figure}

We simulate data in $p=2, 3$ and 4 dimensions. Three different sample sizes $n=500, 1000$ and 2000 are considered. 
For each copula and sample size setting, we replicate the experiment 50 times. 
For one data set generated in a replication, the quality of an estimate $\hat c(\u)$  of the true copula density $c(\u)$ is measured by the mean absolute error ($MAE$) evaluated on an equally spaced $p$-variate grid with 
$M$ 
points on each axis, with left end of the grid value 0.01 and right end of the grid value 0.99 on each axis:
\begin{align*}
MAE = \frac{1}{M^p}\sum_{i_1=1}^M\cdots \sum_{i_p=1}^M |\hat c(u_{i_1}, \cdots, u_{i_p}) - c(u_{i_1}, \cdots, u_{i_p})|,
\end{align*}
where
\begin{align*}
u_{i_j} = 0.01 + (i_j-1)\frac{0.98}{M-1}, \mbox{ for }j=1, \ldots, p.
\end{align*}
The total number of equally spaced grid points inside the unit hypercube $[0,1]^p$ for $MAE$ computation is thus  
equal to $M^p$. 
We use 
$M = 100, 50, 25$ 
corresponding to $p=2, 3, 4$ respectively.    
We report the  
boxplots of the $MAE$ across 50 replications in Figure \ref{CF} through Figure \ref{T5T15N}. The  boxplots for the proposed CFGTN estimator is colored \textcolor{blue}{blue} and labeled by letter `m' on the x-axis. The boxplots for the kernel copula density estimator is colored \textcolor{red}{red} and labeled by letter `k' on the x-axis. The 3 different sample sizes are indicated by the number 1, 2, 3 on the x-axis labels which correspond to sample size $n=500, 1000, 2000$ respectively.

The proposed CFGTN estimator outperforms the kernel copula density estimator, often times by considerable
margins. The mean absolute errors decrease with increased sample sizes for both estimators.  The mean absolute errors under the same model and sample size setting increase when the Kendall's $\tau$ changes from low dependence with value 0.2, to moderate dependence with values 0.4, 0.6, then to high dependence with value 0.8 for both estimators. 

\section{Empirical Applications}  \label{sec:real}

First,  we investigate long term interest rates in four OECD countries - Canada, Greece, France, Italy - from April 1953 to August 2018 (\cite{OECD2018}). The data set includes values of $n=329$ monthly interest rates. 
For each of the 4 univariate marginal distribution, the T-distribution with parameters estimated by maximum likelihood method was found to be adequate. The fitted copula model parameters by CFGTN estimator is listed in Table \ref{table:LTIR-CFGTN} along with their standard errors (SEs) obtained by bootstrapping method. For computing the bootstrap SEs, we drew 200 bootstrap samples. 
For comparison, we compute the kernel copula density estimator (1) \texttt{np} (CV) -  using the quadratic Epanechnikov kernel and optimal bandwidth selected with likelihood cross-validation; (2) \texttt{np} (normal) -  using the normal kernel and optimal bandwidth selected with the normal reference rule-of-thumb. Moreover, the CFGTN copula density estimate is compared with the \texttt{pencopula} - the nonparametric copula density estimate by penalized hierarchical B-splines.  The optimal smoothing parameter $\lambda$ is selected by a simple grid search over 10 equally spaced values from 0.001 to 0.1.     
For the spline dimension $d=3$, the hierarchy order $D=3$, the total CPU time over the ten $\lambda$'s is 898.24 seconds. The CPU time for the selected $\lambda=0.056$ over this grid is 26.11 second. 
For $d=3$, $D=6$, the total CPU time over the ten $\lambda$'s is 9334.79 seconds. The CPU time for the selected $\lambda=0.023$ over this grid is 1395.62 seconds, which is much longer than the time for CFGTN and \texttt{np}. The classical parametric T-copula is also fitted for comparison. The results in Table \ref{table:LTIR} shows that the CFGTN estimator outperforms the other estimators.  

\begin{table}
	\begin{center}
		\caption{Mixture components, estimated proportions, parameter estimates along with the SEs of the copula density estimates by CFGTN for the long term interest rates in Canada, Greece, France, Italy from April 1953 to August 2018}
		\label{table:LTIR-CFGTN}
		\begin{tabular} {c| c  |c}
			\toprule
			\textbf{Component}       &  \textbf{ Proportion (SE)} & \textbf{Copula Parameter (SE)} \\         
			\hline
			Clayton         &   0.1175(0.0461)	& 0.3545(0.4256) \\
			\hline
			Student         &   0.2744(0.0402) & 
$\begin{array}{l}
\hat R_T = 
\begin{bmatrix}
0.9082(0.0653) &               &    \\
0.9488(0.0215) & 0.9517(0.0227) &          \\
0.9424(0.0235) & 0.9399(0.0251) & 0.9964(0.0011) \\
\end{bmatrix}, \\
\hat\nu = 805.1439(212.3835)
\end{array}$	
		\\
			\hline
			Normal 1      &   0.4062(0.0593) & 
$\hat R_1 = 
\begin{bmatrix}
0.7759(0.2423) &               &    \\
0.7394(0.7892) & 0.9205(0.5030) &          \\
0.5683(0.5672) & 0.7614 (0.6738) & 0.8659(0.4152) \\
\end{bmatrix} $
			\\
			\hline 
			Normal 2      &   0.1010(0.0386) & 
$\hat R_2 = 
\begin{bmatrix}
0.3782(0.2162)  &                   &                    \\
-0.6901(0.5899)  & -0.5233(0.4882)  &                    \\
-0.4365(0.6808)  & -0.4907(0.6104)  & 0.8825(0.3975)     \\  
\end{bmatrix}$
            \\
			\hline 
            Normal 3      &   0.1009(0.0346) & 
            $\hat R_3 = 
            \begin{bmatrix}
				0.8011(0.2061)  &                 &                     \\
				0.6278(0.8198)  & 0.6162(0.9661)  &                     \\
				-0.5312(0.6339)  & -0.6025(0.8318)  & 0.1346(0.4439)    \\ 
				            \end{bmatrix} $
		    \\
            \hline	
		\end{tabular}
	\end{center}
\end{table}

As a second example,  we investigate the daily exchange rates of the six currencies to US Dollar:  Euro (EUR), British Pound (GBP), Canadian Dollar (CAD), Swiss Franc (CHF), Japanese Yen (JPY) and Singapore Dollar (SGD) from Jan-03-2000 to May-06-2011 obtained from the Federal Reserve System (\url{https://www.federalreserve.gov/}). 
We model the dependence structure of the log-returns of these six exchange rates by copulas. 
For each of the 6 univariate marginal distribution, the T-distribution with parameters estimated by the maximum likelihood method is used.
We first fit a copula model to the entire data set with $n=2854$ observations using our proposed CFGTN method and compare it with the \texttt{np} (CV) and \texttt{np} (normal). As in \cite{KauermannMeyer2014}, we then divide
the data into time points ante and post the 2008 financial crisis, respectively. As
dividing time point we use September 15th, 2008, the day of the Lehman Brothers
bankruptcy. This leaves us with 2,191 observations prior to the Lehman crisis and 663
observations afterwards. We then fit CFGTN copula models to the two sub-datasets separately, and compare them with the mixtures of Archimedean copulas via simulation-based Bayesian posterior computation as in Table 10 of (\cite{KauermannMeyer2014}) (KM). 
The two copula models based on pre and post Lehman Brothers bankruptcy respectively provide a better fit than the single  copula model based on the entire data set. 
The results in Table \ref{table:EXrate} shows CFGTN estimator outperforms the other estimators.  
The fitted copula model parameters by CFGTN estimator is listed in Table \ref{table:EXrate-CFGTN} along with their standard errors (SEs) obtained by bootstrapping method.
\begin{table}
	\begin{center}
		\caption{Log-likelihood, AICc, CPU Time (in seconds) of the copula density estimates by CFGTN, \texttt{np}, \texttt{pencopula}, T-copula for the long term interest rates in Canada, Greece, France, Italy from April 1953 to August 2018}
		\label{table:LTIR}
		\begin{tabular} {c|c c c}
			\toprule
			\textbf{Method}       &  \textbf{Log-likelihood}  & \textbf{AICc}  &  \textbf{Time}  \\         
			\hline
			CFGTN        &   \textbf{570.11}    &   \textbf{-1073.96}  &  11.46  \\
			\hline
			\texttt{np}(CV) &    432.43    &  -           &  3.38   \\
			\texttt{np}(normal)&    375.95    &  -           &  0.65   \\
			\hline
            \texttt{pencopula} ($d=3$, $D=3$) &   311.41   & -523.13  & 26.11  \\
			\texttt{pencopula} ($d=3$, $D=6$) &   428.93  & -691.32  &   1395.62 \\
			\hline
			T-copula & 509.204 & -1004.057 & 0.661  \\
			\hline
		\end{tabular}
	\end{center}
\end{table}
\begin{table}
	\begin{center}
		\caption{Log-likelihood, AICc, CPU Time (in seconds) of the copula density estimates by CFGTN, \texttt{np},  T-copula, mixtures of Archimedean copulas (\cite{KauermannMeyer2014}) (KM) for the daily Foreign Exchange (FX) rates of EUR, GBP, CAD, CHF, JPY, SGD to US Dollar from Jan-3-2000 to May-6-2011 }
		\label{table:EXrate}
		\begin{tabular} {l| c | c| c  c | c | c }
			\toprule
		        \textbf{Data Set}  &         & \textbf{CFGTN}   &   \textbf{\texttt{np}(CV)}  & \textbf{\texttt{np}(normal)} & \textbf{T-copula}   & \textbf{KM} \\     
		    \hline  
			 FX rate    & Log-like  &  \textbf{5258.40} &       4035.11     &    4216.69         &  5071.87   &    \\
			Jan-3-2000 to & AICc    & \textbf{-10406.67} &                 &                     & -10111.55 &    \\
			May-6-2011    & Time    & 109.67 &         1305.39      &      8.31           &   3.78     &\\ 
			\hline     
			 FX rate    & Log-like  &  \textbf{4189.20}      &   3679.93    &   3331.25          &   4114.51   &  1678.08 \\
            Jan-3-2000 to & AICc    &  \textbf{-8301.02}  &               &                     &  -8196.80   &   -3344.13\\
            Sep-15-2008    & Time   & 94.96        &     650.75    &      4.92           &    2.91    &   \\
             \hline
			 FX rate    & Log-like  & \textbf{1237.64}      &   1154.63    &    1003.99         &    1167.18  &  525.66 \\  
            Sep-15-2008 to & AICc   & \textbf{-2357.52}    &               &                     &  -2301.53 & -1039.07 \\
            May-6-2011 & Time       & 29.26        &      105.52   &      0.53           &    1.25    &          \\
\hline
		\end{tabular}
	\end{center}
\end{table}
\begin{table}
	\begin{center}
		\caption{Mixture components, estimated proportions, parameter estimates along with the SEs of the copula density estimates by CFGTN for the Foreign Exchange (FX) rates data}
		\label{table:EXrate-CFGTN}
		\begin{tabular} {l| c | c | c | c | c | c |c }
	\toprule
	\textbf{Data Set}  &         & \textbf{C}   &   \textbf{F}  &      \textbf{G} & \textbf{T}   & \textbf{N1} & \textbf{N2} \\     
	\hline  
	FX rate            & Prop.   &     0.0160     &   0.0279    &     0.0102     &    0.7306     &    0.1084    &  0.1069  \\
	Jan-3-2000 to      & (SE)    &    (0.0040)    &   (0.0041)  &    (0.0033)   &    (0.0432)    &    (0.0381)  &  (0.0140) \\
	                      \cline{2-8}
	May-6-2011         & Para.   &   0.2060       &    5.6964   &    7.3026     &   $\hat R_T, ~\hat\nu = 7.4175$    & $\hat R_1$      &  $\hat R_2$ \\ 
                       & (SE)    &   (0.0394)     &    (0.5413)    &  (0.1299)  &    (0.4433)     &      omitted              &  omitted \\
	\hline     
	FX rate            & Prop.   &  0.0226        &    0.0290    &              &    0.7875    &      0.1509                 &  \\
	Jan-3-2000 to      & (SE)    &   (0.0013)     &   (0.0016)     &            &  (0.0084)   &    (0.0048)    &  \\
 	              \cline{2-7}
	Sep-15-2008        & Para.    &   0.8809       &     5.2872     &           &   $\hat R_T, \hat\nu =  12.5141 $    &   $\hat R_1$             & \\
	                   & (SE)    &   (0.0576)       &   (0.2411)   &            &   (0.0340)    &       omitted               &  \\
	\hline
	FX rate            & Prop.   &   0.0196     &     0.0375    &     0.0419    &    0.5942     &      0.2536      &  0.0533 \\    
	Sep-15-2008 to     & (SE)    &   (0.0122)    &    (0.0221)    &    (0.0249)   &   (0.0616)    &    (0.0537)    &  (0.0146) \\
                	\cline{2-8}
	May-6-2011          & Para.    &  0.0222      &    5.9199     &     1.8937  &    $\hat R_T, \hat\nu = 6.3290$    &        $\hat R_1$         &   $\hat R_2$ \\
	                    & (SE)    &    (0.3760)    &    (0.4984)    &   (0.4146)   &    (1.2373)    &        omitted               &   omitted   \\
	\hline
       \end{tabular}
	\end{center}
\end{table}
\section{Concluding remarks} \label{sec:con}

We presented a finite mixture of Clayton, Frank, Gumbel, T, and normal copula components model. 
The model parameters are estimated by the interior-point algorithm for the resulting constrained maximum likelihood estimation problem, 
where the gradient of the objective function is not required. 

The general purpose MATLAB function \texttt{fmincon()} works well for data sets in moderate dimensions such as 3, 4, 5, 6.  
A custom designed code which utilizes the gradient of the objective function may bring up the speed. 
  
The theoretical questions such as the consistency and convergence rate
of the estimator wait to be investigated. For a probability density modeled by a finite mixture of densities from the same family,  
\cite{Leroux1992} discussed the use of AIC and BIC for order selection and proved their consistency.  
The extension of this result to the case of a heterogeneous copula density mixture would be interesting and challenging.


\if0\blind
{
\section*{Acknowledgment}
We thank the anonymous referees for insightful and constructive comments which have helped us to significantly improve the paper. 

We used matlab functions \texttt{mvcoprnd()}  and \texttt{copulaparam()} provided by Robert Kopocinski in matlab central file exchange [\cite{Kopocinski2007}]. 

We used multivariate Archimedean copula matlab functions provided by Martin Scavnicky in github [\cite{Kopocinski2012}]. 
} 

\fi


\bibliographystyle{model2-names}
\bibliography{Qu_JournalsFull,Qu_Copula_References_2019,Qu_Mutual-Information_References_2015,Qu_Statistics_References_2019,Qu_Optimization_References_2019,Qu_Model-Selection_References_2018}	
\end{document}